\documentclass[12pt,preprint]{aastex}

\slugcomment{}

\newcommand{\gI}{$g\!-\!I$}
\shorttitle{Globular Cluster Populations in AM 0139-655}
\shortauthors{Maybhate et al.}

\begin{document}


\title{Evidence for Three Subpopulations of Globular Clusters in the Early-Type Post-Starburst Shell Galaxy AM 0139-655\altaffilmark{1}}


\author{Aparna Maybhate\altaffilmark{2}, Paul Goudfrooij\altaffilmark{2}, Fran\c cois Schweizer\altaffilmark{3}, Thomas Puzia\altaffilmark{4}, David Carter\altaffilmark{5}}

\altaffiltext{1}{Based on observations with the NASA/ESA {\it Hubble Space Telescope}, obtained at the Space Telescope Science Institute, which is operated by the Association of Universities for Research in Astronomy, Inc., under NASA contract NAS5-26555}
\altaffiltext{2}{Space Telescope Science Institute, Baltimore, MD 21218; maybhate@stsci.edu, goudfroo@stsci.edu}
\altaffiltext{3}{Carnegie Observatories, 813 Santa Barbara Street, Pasadena, CA 91101; schweizer@ociw.edu}
\altaffiltext{4}{Herzberg Institute of Astrophysics, 5071 West Saanich Road, Victoria, BC V9E 2E7, Canada; puziat@nrc.ca}
\altaffiltext{5}{Astrophysics Research Institute, Liverpool John Moores University, 12 Quays House, Egerton Wharf, Birkenhead, CH41 1 LD, United Kingdom; dxc@astro.livjm.ac.uk}

\begin{abstract}
We present deep {\it Hubble Space Telescope} ACS images of the post-starburst shell
 galaxy AM 0139-655. We find evidence for the presence of three distinct
globular cluster subpopulations associated with this galaxy: a centrally
concentrated young population ($\sim$\ 0.4 Gyr), an intermediate age population
($\sim$\ 1 Gyr) and an old, metal-poor population similar to that seen around normal galaxies.
The \gI\ color distribution of the clusters is bimodal with peaks
at 0.85 and 1.35. The redder peak at \gI\ = 1.35 is consistent with the predicted color
for an old metal-poor population.
The clusters associated with the peak at \gI\ = 0.85 are centrally concentrated
and interpreted as a younger and more metal-rich population. We suggest that
these clusters have an age of $\sim$\ 0.4 Gyr and solar metallicity based on a
comparison with population synthesis models.
The luminosity function of these ``blue" clusters is well represented by a power law,
$\phi(L)dL \propto L^{-1.8} dL$. Interestingly, the brightest shell
associated with the galaxy harbors some of the youngest clusters observed.
This seems to indicate that the same merger event was responsible for the
formation of both the shells and the young clusters.
The red part of the color distribution contains several very bright clusters,
which are not expected for an old, metal-poor population. Furthermore, the
luminosity function of the ``red" GCs cannot be fit well by either a single
gaussian or a single power law. A composite (gaussian + power law) fit to
the luminosity function of the red clusters yields both a low rms and very
plausible properties for an old population (with gaussian distribution) plus
an intermediate-age population (with power-law distribution) of GCs.
Hence, we suggest that the red clusters in AM 0139-655 consist of two
distinct GC subpopulations, one being an old, metal-poor population as seen in
normal galaxies and one having formed during a recent dissipative galaxy
merger (likely the same event that formed the $\sim$\, 0.4 Gyr old clusters).
\end{abstract}



\section{Introduction}
Globular clusters (GCs) are very useful probes of the dynamical and chemical assembly
histories of galaxies.
Many globular cluster systems in ``normal" giant elliptical galaxies show
a bimodal color distribution, providing clear evidence for the presence of a
``second event" during the formation of
these systems \citep{ash98}. Spectroscopic studies
have revealed that both the blue and red GC populations are typically old
($\ge$ 8 Gyr, \citep{forbes01,cbc03,puzia05}) implying
that the bimodality is mainly due to a difference in their metallicity.
The merger scenario
of \cite{ash92} predicted the presence of this bimodal distribution. However,
different theories were suggested by various authors to explain the nature
of the second event. The merger model \citep{schweizer87,ash92}, predicts that the metal-rich clusters
are formed during major mergers of gas-rich galaxies. The ``multi-phase
collapse" model \citep{forbes} implies that they form during a secondary
collapse phase of the galaxy, without invoking a merger event. The accretion
scenario \citep{fmm,cmw} proposes that every galaxy is born with a GC
system that has a median color (metallicity) according to the color-magnitude
relation of galaxies \citep{terlevich}. The bimodality then arises due to the accretion
of low-mass galaxies with metal-poor GCs by a larger galaxy with its
metal-rich GCs.

Intermediate-age GCs have been found in several galaxies \citep{gmk,gnew, puzia02,
kundu05,chandar}.
A few of these galaxies show features like shells or a disturbed morphology,
indicating a recent merger event. Shells are sharply defined, arc-like features, mostly found around elliptical
galaxies \citep{malin}, though a few (Arp 215, Arp 227) have been found
around disk galaxies \citep{arp}.
Two classes of models have been proposed to account for the shells. The
internal models \citep{fabian,wc} suggest that shell stars form because of
star formation within a giant shell of shocked interstellar gas.
On the other hand, the interaction models suggest an external origin for the shell structure and
interpret it as arising due to a merger. 
When an elliptical galaxy accretes a small
companion galaxy, some vestiges of the event survive for a Gyr or longer after the
merger in the form of ripples and shells \citep{schweizer1980,quinn1984}.
The morphology and alignment of the resulting shell system depends on the form of the potential of the primary galaxy, the orbital parameters of the collision and the nature of the companion \citep{hq88,hq89,dc86}.
Typical dynamical ages of shells are
0.5 - 2 $\times 10^9$ years \citep{nulsen89,hq87}. These shells, made up of stars, have
been frequently observed around ellipticals. \cite{malin} find 17$\%$ and
\cite{sf} find 44$\%$ of field ellipticals to be surrounded by shells.
The presence of shells thus seems to be a common outcome of interactions of
galaxies.

To test the merger scenario for GC formation, it is important to study galaxies that show obvious signs of
a merger event in the form of shells, ripples, and some signature of having
hosted a strong burst of star formation.
One way of attempting this is to study E+A galaxies featuring shells. E+A
galaxies were originally defined by the combination of an early-type galaxy
morphology and
the presence of a so-called ``post-starburst''
spectrum containing strong Balmer absorption lines (indicative of A-type stars)
without strong [O\,{\sc
ii}] or H$\alpha$ emission lines \citep{dressler83}\footnote{ Subsequent
studies revealed the presence of a disk component in many such galaxies,
and many papers now use the term k+a, where the k stands for the
spectral type of an old stellar population \citep[e.g.,][]{franx93}.}.
The existence of strong Balmer absorption
lines indicates that these galaxies have experienced star formation
around $\sim 0.3 - 2$ Gyr ago while the absence of emission lines 
indicates that star formation must have ceased shortly thereafter.
A large fraction (75$\%$) of nearby E+A galaxies are located in the field \citep{zab96}. Several of these show tidal features indicative of galaxy-galaxy interactions and mergers as the probable mechanism responsible for their formation.

Long-slit spectroscopy of E+A galaxies \citep{norton} revealed that the young stellar populations are typically more centrally concentrated than the old population, but are not always confined to the inner 1 kpc.
\cite{sch} discovered atomic hydrogen associated with the shells of NGC 5128.
Subsequently, CO molecules were found in two shells surrounding this elliptical
 providing evidence that the
molecular gas in mergers may not only fall into the nucleus but can also be spread out to large galactocentric radii \citep{char}.
\citet{carter} found evidence for recent nuclear star formation in 20 out of 100 shell galaxies in the
\cite{malin} catalog, suggesting that the process of
shell formation may involve some accretion or merger of gas-rich material
which would account for the nuclear star formation.

Numerical simulations of the sinking of a gas-rich
satellite disk galaxy into a large elliptical galaxy by \citet{kn}
show that the stellar particles from the disk stars of the satellite
start making a clear shell structure after the first collision with the
elliptical in case of a radial or a retrograde merger. This shell system lasts
for $\sim$ 1 Gyr or more. They also find an abrupt reduction in the star formation rate
around the time of first passage of the satellite through the center
of the elliptical due to the sudden decrease in gas density caused by the
gravitational scattering by the elliptical potential.

The
study of GCs in such galaxies should yield important information regarding
the nature of the E+A phenomenon. If numerous intermediate-age GCs with
ages consistent with those of the post-starburst population in the
nucleus were to be found, we could be fairly confident that the A
stars were formed in a relatively vigorous and distinct star formation
event, likely associated with a dissipative galaxy merger. Conversely,
an absence of such GCs would indicate that no strong star formation was
associated with the event that caused star formation to end $\sim$\,0.5 to
2 Gyr ago. The latter would be consistent with a scenario where gas is
removed from the galaxy, e.g.\ by ram pressure stripping. This paper
focuses on a study of the globular cluster system of the early-type shell
galaxy AM 0139-655.

\section{AM 0139-655}
\objectname{AM 0139-655} (PGC 6240) is a giant early-type galaxy with
bright shells that are irregularly distributed around the main body
\citep{malin}.
The nuclear spectrum of AM 0139-655 shows evidence for a post-starburst
(E+A) spectrum with very strong Balmer absorption lines (suggesting the
presence of a substantial fraction of A-type stars)
but no significant [O\,{\sc ii}] emission \citep{carter}.
AM 0139-655 has a large H$\delta$ equivalent width (13.6 $\pm$ 1.0 \AA) and
an active nucleus.
Since AM 0139-655 shows both shell structure (indicative of a merger)
and a post-starburst spectrum (signaling a burst of star formation and hence
possible
globular cluster formation in the recent past),
it is a good  candidate to explore whether any relations exist between its
globular cluster system, its morphology and the merger event that gave rise
to the shells.
In this paper, we present a detailed photometric study of the GCs of AM 0139-655.
The general properties of AM 0139-655 are listed in Table~\ref{tbl-1}.

\section{Observations and Data Reduction}

\begin{figure}
\plotone{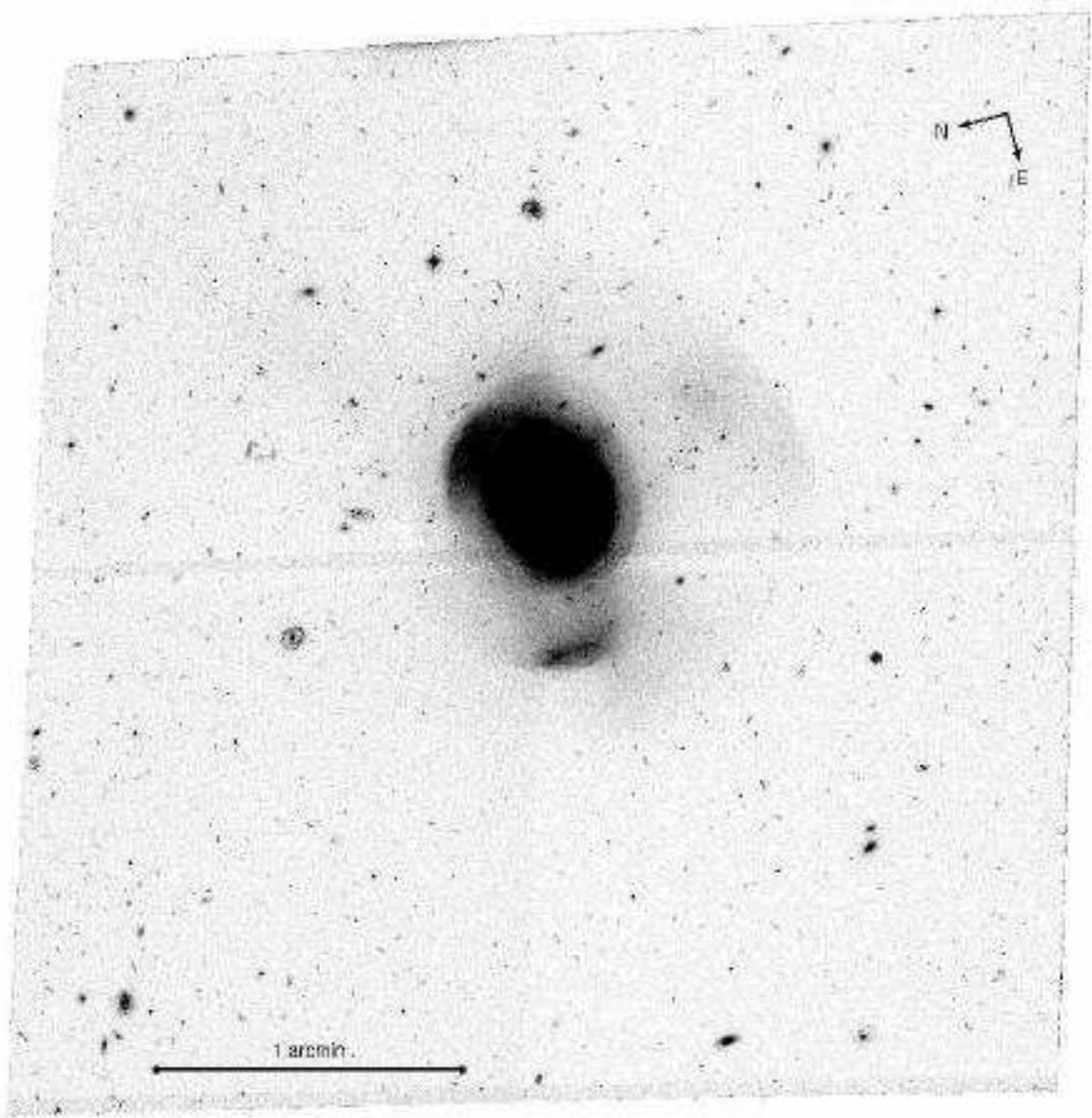}
\caption{ACS image of AM 0139-655 in F475W \label{img}}
\end{figure}

\objectname{AM 0139-655} was observed with {\it HST} on July 16 and 17, 2004,
as part of General Observer program 10227, using
the Wide Field Channel (WFC) of ACS. It was observed for a total of
6 orbits in the continuous viewing zone (CVZ) of HST with total integration times
of 21440 s in the F475W and 7300 s in the F814W filter.
The observations consisted of several long exposures at different dither
positions as well as
a few short ones to prevent saturation of the galaxy center.
The data were processed with the ACS on-the-fly pipeline which
included dark and bias subtraction and flat fielding. Each flat-fielded
image was carefully
checked for satellite trails and saturated pixels in the central region
of the galaxy. These pixels were flagged and masked out. The individual images
in each band were co-added using the PyRAF task MULTIDRIZZLE \citep{kok}. This procedure
removed all the cosmic rays and corrected for geometric distortion.
The combination
of the short and long exposure images enabled us to get good resulting
multidrizzled images with unsaturated centers. The final combined F475W image
is shown in Fig.~\ref{img}.



\section{Photometry}
The search for sources was performed on an image made from the summed F475W+F814W
image in the following manner.
The F475W and the F814W images from multidrizzle were added
together. To improve the detection and the accuracy of the photometry, especially for the faint sources embedded within the bright galaxy background, the
underlying galaxy light was first modeled. Extensive experimentation showed that the application of a circular median filter gave the best results. 
Object detection was
done by using the DAOFIND routine in the DAOPHOT package \citep{stet} on
an image obtained by dividing the combined image by the square root of
the median filtered image (thus providing uniform shot noise characteristics
over the whole frame). 
The advantage in using the {\it g+I} image rather than the individual images
for source detection is the greater depth reached by the combined image. Also,
the photometric zero-point of {\it g+I} is significantly less dependent on the
color of the source than the individual {\it g} or {\it I} images, which helps
during the completeness correction (see Goudfrooij et al. 2007 for a detailed discussion).
The detection threshold was set at 4$\sigma$ above
the background. Typical half-light radii of globular clusters
are in the range 1- 20 pc \citep{kundu,vanden,jordan}. At the distance of
AM 0139-655, the scale is 25 pc per ACS$/$WFC pixel.
Thus, we expect the globular clusters to be nearly unresolved point sources on the image.

Smooth, median-filtered model images  were created in each of the two bands from the
corresponding multidrizzle combined images in a manner similar to the one
described above. The model was subtracted from
the corresponding original images to get a residual image in which the galaxy light has been
removed to a great extent and the compact objects are clearly visible.
To confirm that the subtraction
procedure did not alter the error statistics of the photometry, we compared the
magnitudes obtained from the subtracted and the original images for a few bright
sources in the frame. The differences between the two methods were well below the
photometric errors.
Aperture photometry was carried out on these residual images through an
aperture of 3 pixel radius (0\farcs15). In order to discard obvious spurious
detections (e.g. edges of the image), objects with magnitude errors greater than 0.4 mag
were rejected at this point.
Aperture corrections from
3 pixel (0\farcs15) to 10 pixel (0\farcs5) radius were determined using a
few bright point sources in each band. The corrections from 0\arcsec .5 to
infinity were taken from \citet{siri05}. The total aperture corrections
were 0.470$\pm$0.009 and 0.517$\pm$0.010 in F475W and F814W respectively.
The transformation from the instrumental magnitudes (STMAG) to
SDSS {\it g}- and Cousins {\it I}-band was done using the SYNPHOT package in STSDAS.
Synthetic spectra of stellar types ranging from B0V to K3III from the
\citet{bru} library were convolved with the SDSS {\it g} and the Cousins {\it I}
filter to obtain magnitudes relative to Vega, and with the F475W and
the F814W filters to yield magnitudes in the STMAG system. Fitting polynomials
to get the relation between the two systems resulted in the following
transformations:

\begin{eqnarray}
g & = & F475W + (0.4530\pm0.0006)\nonumber \\
      &&+ (0.039\pm0.001)(F475W-F814W)\nonumber \\
     &&  + (0.0010\pm0.0003)(F475W-F814W)^2 \nonumber \\
      && - (0.0040\pm0.0003)(F475W-F814W)^3 \nonumber \\
I & = & F814W - (1.258\pm0.002)\nonumber \\
      &&+ (0.015\pm0.002)(F475W-F814W)\nonumber \\
       &&- (0.002\pm0.003)(F475W-F814W)^2\nonumber\\
       && - (0.001\pm0.001)(F475W-F814W)^3
\end{eqnarray}

\subsection{Globular Cluster Candidate Selection}

The photometry lists with the magnitudes in {\it g} and {\it I}
obtained in the preceding section also included foreground stars and
extended objects, in addition to cluster candidates. To remove these contaminating sources
we applied two selection criteria based on the FWHM and the compactness
of each source.
We obtained the direct FWHM of each of the sources in the photometry
list in each band using the task IMEXAM within IRAF\footnote{IRAF is distributed by the National Optical Astronomy Observatories,
which are operated by the Association of Universities for Research
in Astronomy, Inc., under cooperative agreement with the National
Science Foundation.} in batch mode. Since the FWHM of
bright stars in our frame is $\sim$2.0 pixels and we expect all the cluster
candidates to be nearly unresolved point sources, we discarded all sources
that did not satisfy the criterion 1.5 $\leq$ FWHM $\leq$ 3.0 pixels. We also
added a compactness criterion wherein only sources that
had the difference in the magnitudes between aperture radii of 2 pixels and 5 pixels
lie between 0.2 mag and 0.6 mag were retained. This ensured (to a great
extent) that the contamination from background galaxies was
minimized.
To study the radial extent and
distribution of the GC candidates so obtained, they were binned into annuli
400 pixels wide, starting from the center of the galaxy out to the corners of the
ACS frame. An effective area was calculated for each annulus namely the total
area of the pixels that lie within the annulus which had a non-zero value of
intensity. This ensured that the masked-out regions like the chip gap were not
considered while computing the area since these areas were also excluded
during cluster finding.
We plot the logarithm of the number of sources per kpc$^2$ as a
function of the projected distance from the center of the galaxy and find
that it reaches an asymptotic value at a projected radius of about 40 kpc
(see
Fig.~\ref{figsrc}). Given the large distance of the galaxy and hence, the small
angular size of AM 0139-655, it was possible to use the outer parts of the
image to estimate the contamination due to compact background sources.

\begin{figure}
\plotone{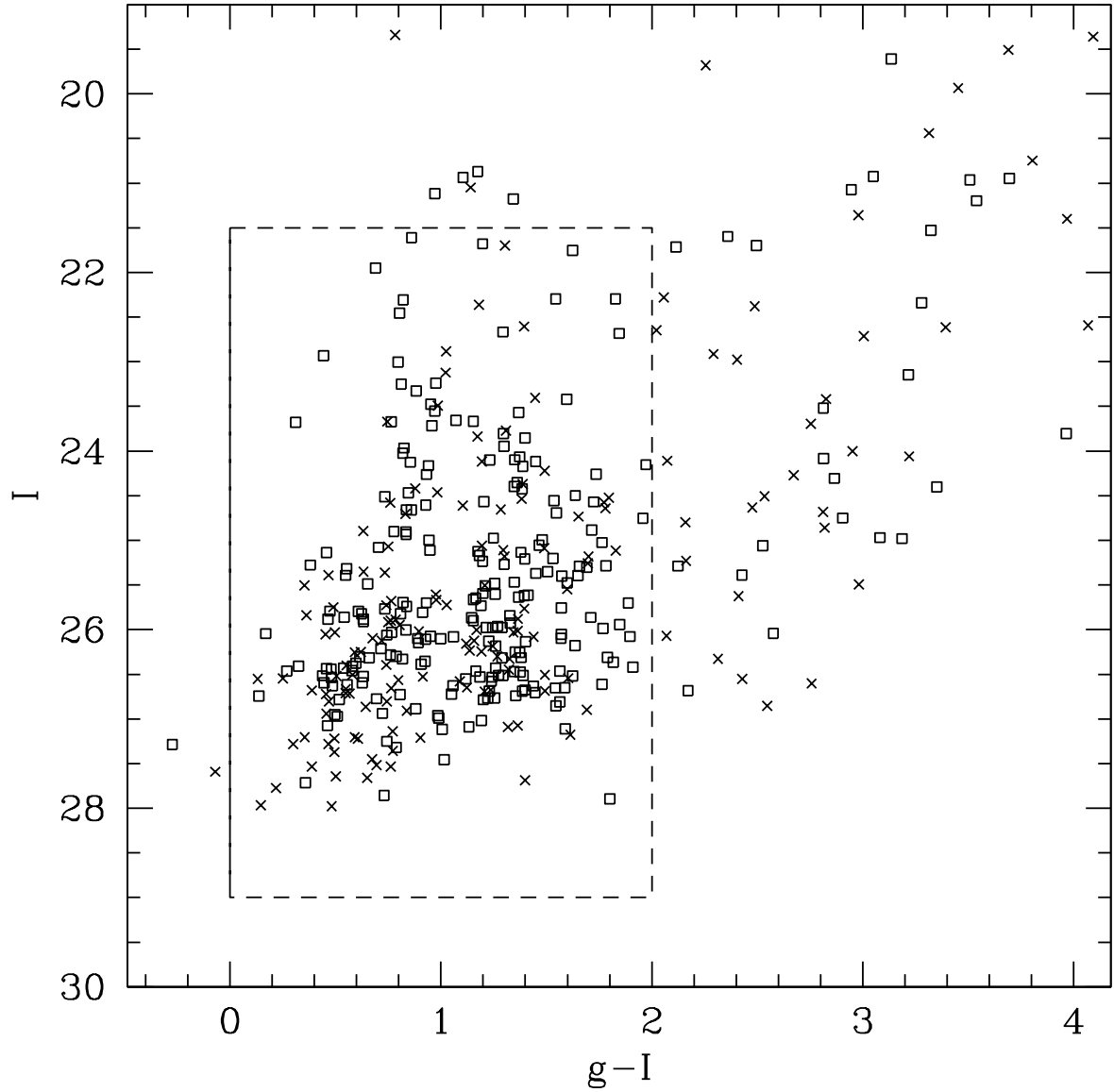}
\caption{{\it g} versus \gI\ color-magnitude diagram for AM 0139-655 globular 
cluster candidates. The open squares denote GC candidates present within the
inner 40 kpc and the crosses denote those present beyond 40 kpc. The dotted lines
delineates the region of GC selection. \label{figgi}}
\end{figure}

From the \citet{burstein} model, we find the Milky-Way foreground extinction $A_V$ = 0.00.
The resulting {\it g} versus \gI\ color-magnitude diagram for the candidate
globular clusters is shown in Fig.~\ref{figgi}. The rectangle delineates
the sources that meet our criteria for globular cluster candidates.
Using the GALEV simple stellar population (SSP) models \citep{galev}, we determine the color
range of 0 $\leq$ \gI\ $\leq$ 2.0 to include all clusters over the
range of ages from $10^{7}$ to $10^{10}$ years and metallicity z values from 0.0004 to 0.05 (where z = 0.02 represents solar metallicity).
All sources with colors
redder than 2.0 are expected to be late-type foreground stars or red
background galaxies and are
rejected in the subsequent analysis. 
The final list, thus obtained, contains a total of 217 globular cluster
candidates. Typical photometric errors in {\it g} were 0.01, 0.03, 0.15 and $>$0.2 mag
for magnitude intervals 22-24 mag, 24-26 mag, 26-28.75 mag and fainter than 28.75 mag respectively. The brightest fifty GC candidates are listed in Table~\ref{tbl-2}.

\subsection{Completeness Corrections}
\begin{figure}
\plotone{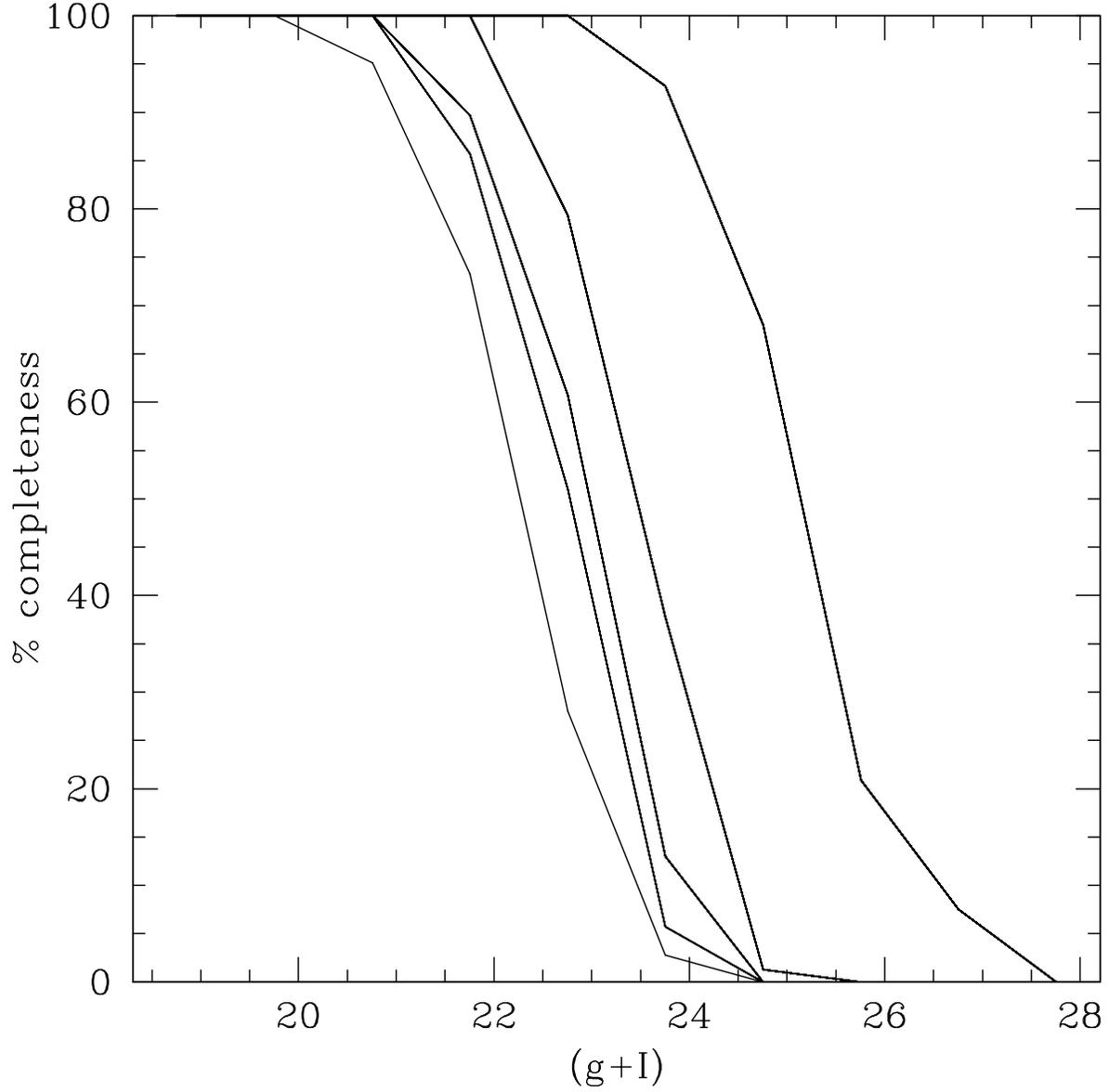}
\caption{Completeness correction curves for the combined g and I band for five background levels. From left to right: 2000, 980, 480, 200, 120 $e^{-} $ per pixel for an effective {\it g+I} exposure time of 1822 s.}
\label{complete}
\end{figure}

Completeness tests were performed on the combined {\it g+I} image, i.e., the one
used for detection of cluster candidates. Artificial objects were added in
batches of 100 for
five background levels and several magnitude intervals. The background
levels were computed from the {\it g+I} image in the following manner.
The counts in the outermost parts of the image, free from any galaxy
light contamination were taken as Background level \#1. The counts in the
innermost region of the galaxy where clusters are detected were taken as
Background level \#5. This difference between the faintest and the brightest background
levels was then divided into five equal
intervals in log space to get the five mean background levels.
The radial intensity profile
of the artificial input globular clusters was determined by fitting PSFs to
the cluster candidates in the original image.
The image was divided by the square root of a median-filtered version of itself and source detection was performed on the resultant image, with the same
procedure as used for the initial source finding. Further, the ranges of
the FWHM, compactness and 
maximum permissible error in the derived magnitudes were also selected
in exactly the same manner as that for the actual source photometry.
The resultant completeness curves are
shown in Fig.~\ref{complete}.

\section{\gI\ Color Bimodality}
To analyze the color distribution of globular clusters in
AM 0139-655, we start by computing the non-parametric Epanechnikov-kernel
probability density function ${\cal P}_{\rm obj}$ of all objects inside
the analysis radius of 40 kpc \citep{silverman86}. Given the photometric
completeness of our data ${\cal P}_{\rm obj}$ is the closest
approximation
to the real color distribution of globular clusters. The width of the
adaptive kernel linearly depends on the statistical color uncertainty of
every individual object, so that the
initial noise statistics of the color measurements is translated into
the
final probability density estimate. To account for uncertainties due to
the limited sample size, we bootstrapped this process 100 times and derive
90\% confidence limits of the resulting probability density function.

\begin{figure}
\plotone{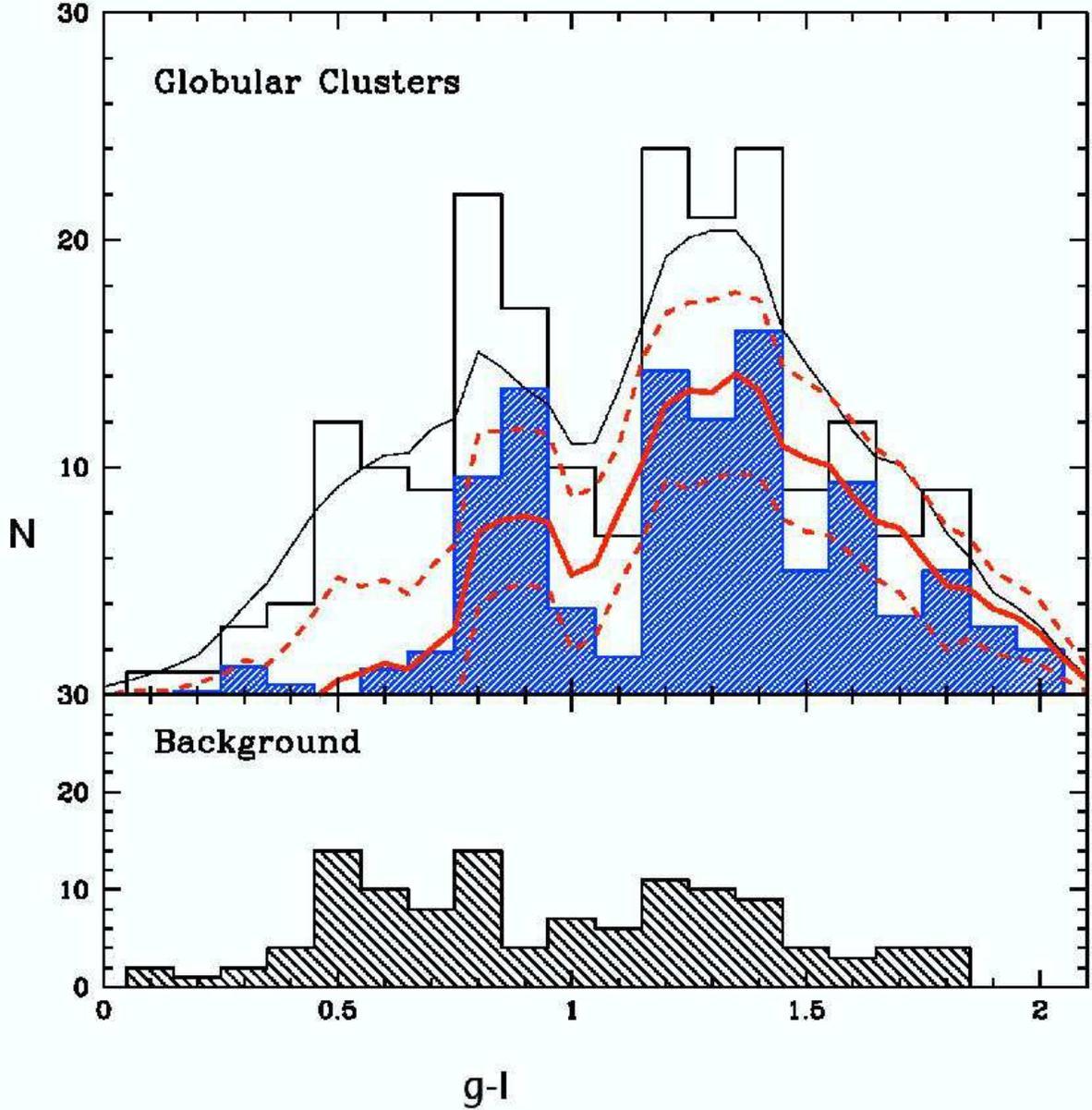}
\caption{GC color distribution for AM 0139-655. The open histogram in the top panel represents the observed color distribution and the hatched histogram represents the distribution after subtraction of the background. The background distribution is shown in the lower panel. The solid red line is a non-parametric probability density estimate using an adaptive Epanechnikov kernel of the background corrected GC color distribution. The dashed red lines mark the bootstrapped 90$\%$ confidence limits. The solid black line represents the probability density estimate of the uncorrected distribution. Two peaks are clearly seen at \gI\ = 0.85 and \gI\ = 1.35. Note that the distribution of red clusters has a broad tail. There is a very faint signature of the redder, metal-rich peak that is seen in normal ellipticals at \gI\ $\approx$ 1.8. This suggests that one or both of the progenitors had only a small bulge.}
\label{gi}
\end{figure}

With the same technique we also computed the probability density function
of the
background field ${\cal P}_{\rm bkg}$ derived from the detections. We then
scaled the background density estimate to
the object function ${\cal P}_{\rm obj}$ by area. We note that the error
associated with the scaling of the function is included in our bootstrap
routine and propagated to the uncertainty of the final density
estimate. We finally derive the color distribution of globular cluster
candidates by statistical subtraction:
\begin{equation}
{\cal P}_{\rm GCC} = {\cal P}_{\rm obj}-A{\cal P}_{\rm bkg}
\end{equation}
where $A$ is the ratio of areas encompassed by the object versus background
regions.
Figure~\ref{gi} shows the background corrected function. The final color
distribution appears clearly extended and, taking the 90\% error margins
into account, bimodal with peaks at \gI\ = 0.85 and \gI\ = 1.35. A rather
extensive red wing to the red peak is present, which will be discussed in
Sect. 7. KMM tests \citep{mb88,abz} were also run on the background
subtracted \gI\ color distribution. The KMM test for two components resulted in
peak values consistent with those found using the Epanechnikov kernel. The test
for three components resulted in a third peak with \gI\ = 1.45 but without any
clusters assigned to it.

Although we do not know the effects of possible internal extinction on individual
GCs, the fact that the blue peak of the GC color distribution found in AM 0139-655
at \gI\ = 1.35 is consistent with $V-I$ = 0.95 (for 14 Gyr old, metal-poor clusters
from SSP models)
which agrees with the blue peak of GCs in the Milky-Way indicates that on average the internal
visual extinction is very small. At the same time, this also supports our assumption that the Milky-Way
foreground extinction is near zero, as predicted by the \citet{burstein} model.

\subsection{Spatial Distribution of the Blue and Red Subpopulations}
As shown in the preceding section, the color distribution of the clusters
has a bimodal form with peaks near \gI\ = 0.85  and 1.35. The spatial
distributions of the two subpopulations were derived by
dividing the clusters into two subpopulations based on their colors.
Clusters which lie in the range 0 $\leq$ \gI $\leq$ 0.95
were designated as the blue population and those within 1.15 $\leq$ \gI$\leq$ 2.0
were designated as the red population.
The region 0.95 $<$ \gI\ $<$ 1.15 was avoided to
minimize mixing of the two populations.
This yielded 84 blue clusters, 114 red clusters and 19 clusters that lie
in the region of mixing.

\begin{figure}
\plotone{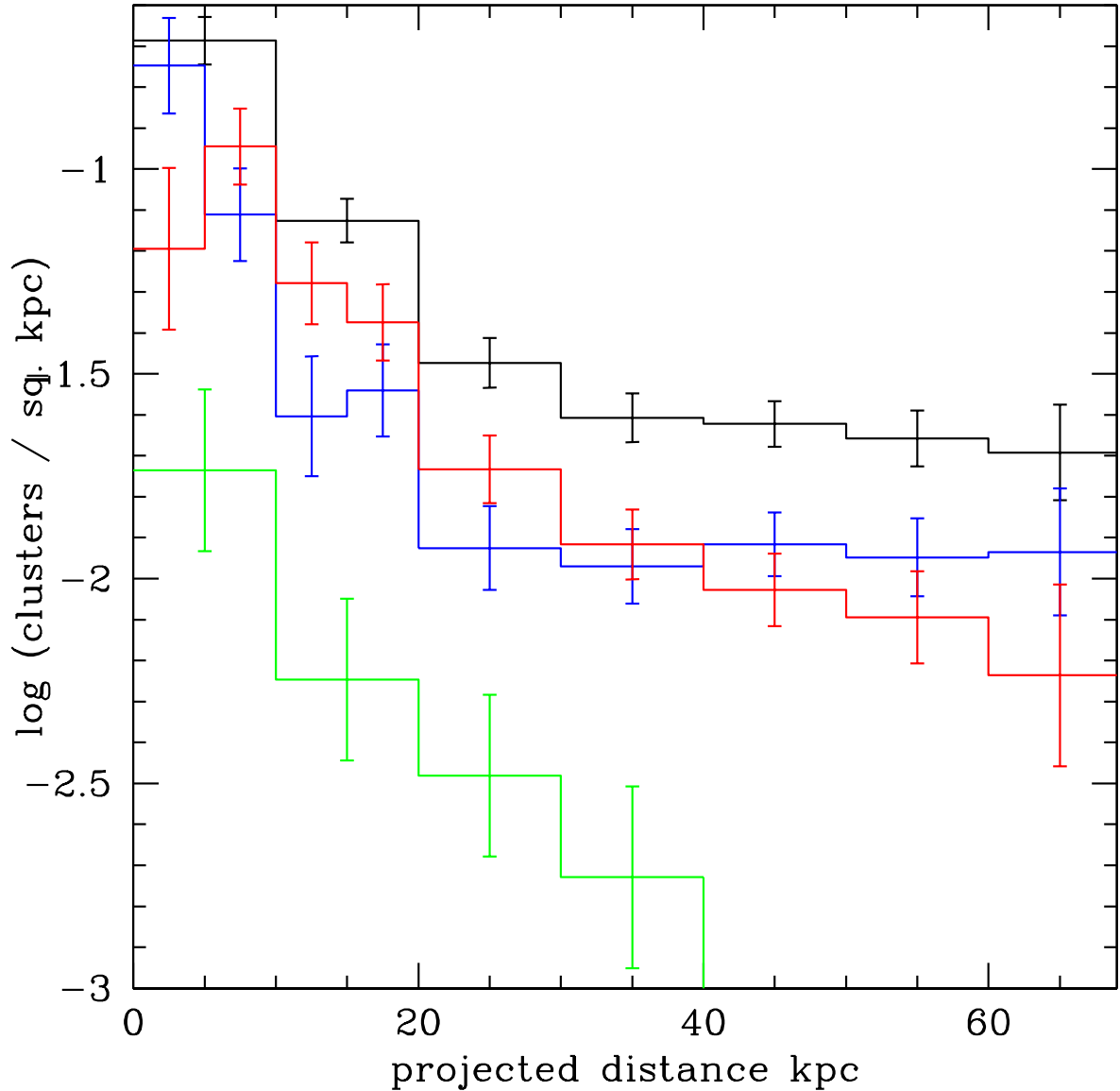}
\caption{Surface density of the globular cluster candidates as a function of the
 distance from the galaxy center. The black solid line shows all the clusters,
the blue line shows the distribution of the blue clusters and the red
line shows the distribution of the red clusters. The surface density of clusters lying in the color
range 0.95 $<$ \gI\ $<$ 1.15 (between the blue and the red clusters) is shown by the green line.
The bin width is 10 kpc
except for the inner 20 kpc for the blue and the red clusters where it is 5 kpc.
The blue clusters seem to be more centrally concentrated as compared to the red clusters. \label{figsrc}}.
\end{figure}

Figure~\ref{figsrc} shows the radial distribution of the blue and red samples.
The bins were chosen in such a way that there were at least five clusters in
each bin in the region within 40 kpc. It
is clear from the figure that the blue clusters are more concentrated
towards the center and their surface density falls as we go away from the
center. On the other hand, the surface density of the red clusters is somewhat
low in the center and may peak between 5-10 kpc from the center of the galaxy.
The surface density distribution of the 19 GCs that lie in the intermediate
color range is also shown in Fig.~\ref{figsrc}. This distribution appears to
peak at the center. Adding all these clusters to the blue clusters would
not change the trend for the blue clusters while adding them to the red
clusters would tend to somewhat dilute the effect in the central regions.

We used the SSP models described in \citet{galev} to estimate the ages of
the two populations. These models include continuum and line emission and take
into account the important contribution of gaseous emission to broad-band
fluxes and their strong metallicity dependence during the very early
evolutionary stages of star clusters. The top panel of Fig.~\ref{time} shows the dependence of \gI\ color on the age of the cluster. From the figure, we see
that the red peak in the color distribution
at \gI\ = 1.35 can be explained by clusters of age 14 Gyr and metallicity
Z= 0.0004 which agrees very well with the colors expected for the old,
metal-poor globular cluster population that forms the blue peak in normal
ellipticals or spirals. The blue peak of the color distribution at \gI\ = 0.85
can be explained by a younger population of clusters
of solar metallicity with an age $\sim$ 0.4 Gyr. Note that this age
estimate is fairly independent of the metallicity: See Fig.~\ref{time}.
This is consistent with
the E+A nature of AM 0139-655.
We assume solar
metallicity for these clusters since several spectroscopic studies of luminous
intermediate-age globular clusters in a number of merger remnants have shown that
they have near-solar metallicities (NGC 7252 \citep{ss}; NGC 3921 \citep{ssb}; NGC 1316 \citep{gmk} and NGC 5128 \citep{pff}).
{\it HST} imaging studies of the same merger remnants
show that these young clusters
are more concentrated towards the galaxy center than the old metal-poor
clusters \citep{mwsf,smwf,gmk,pff}, as predicted by \citet{ash92} who modeled the formation of
clusters in a dissipative galaxy merger.

\begin{figure}
\plotone{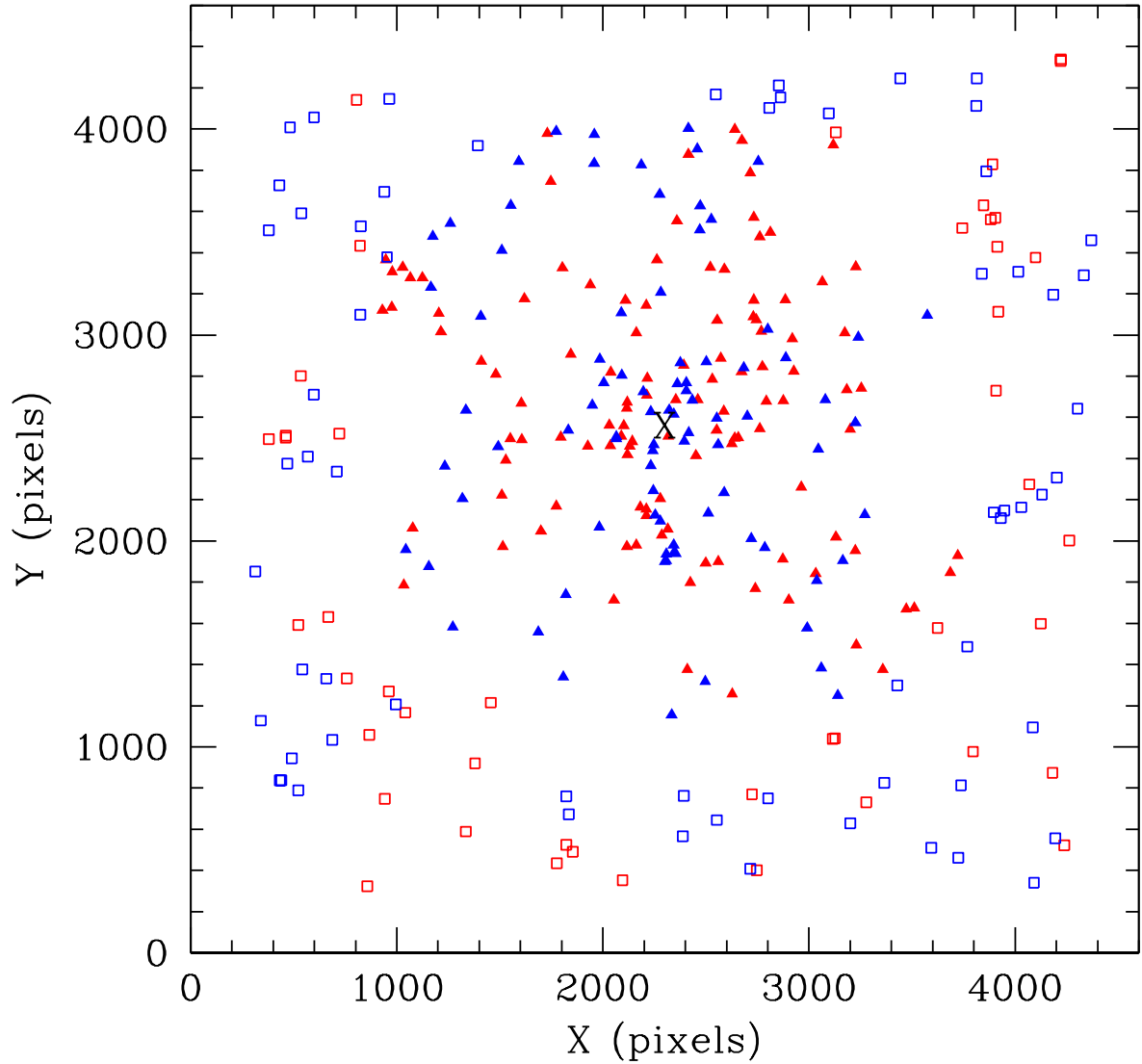}
\caption{The projected spatial distribution of the clusters on the ACS image. ``X" denotes the galaxy center.
The red and blue filled triangles stand for the red and the blue clusters,
respectively. The open red and blue squares are the red and blue objects
in the background region where the surface number density of clusters is approximately
 constant. \label{figredblue}}
\end{figure}

\begin{figure}
\plotone{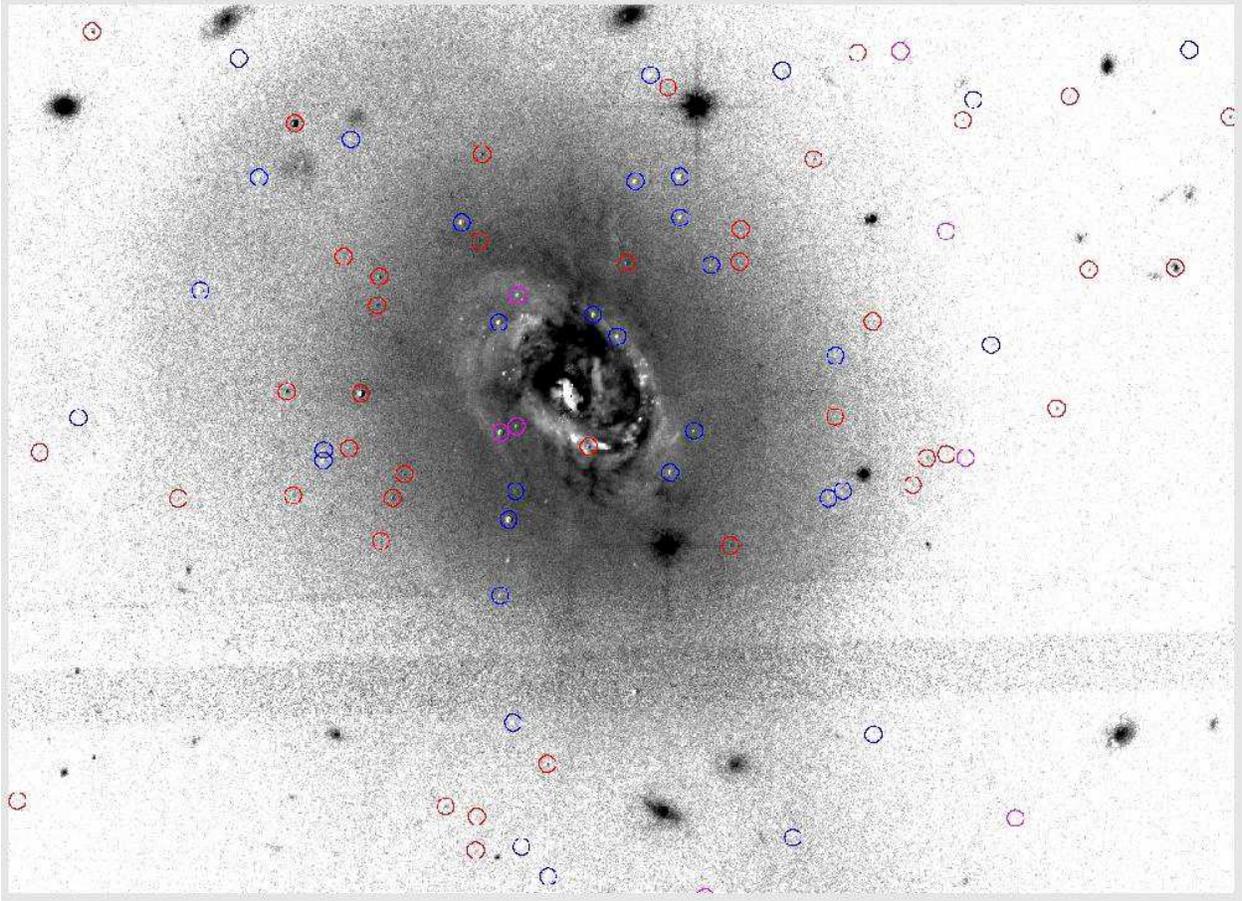}
\caption{\gI\ image of the inner region of AM 0139-655 showing the locations of the globular cluster candidates in the inner region. The dark regions denote redder colors, and the light regions denote bluer colors. The blue, red and magenta circles represent blue clusters, red clusters and clusters with \gI\ between 0.95 and 1.15. The bright region in the center is the nucleus of AM 0139-655. Note the presence of several clusters along the blue ring surrounding the center. \label{ring}}
\end{figure}

\begin{figure}
\plotone{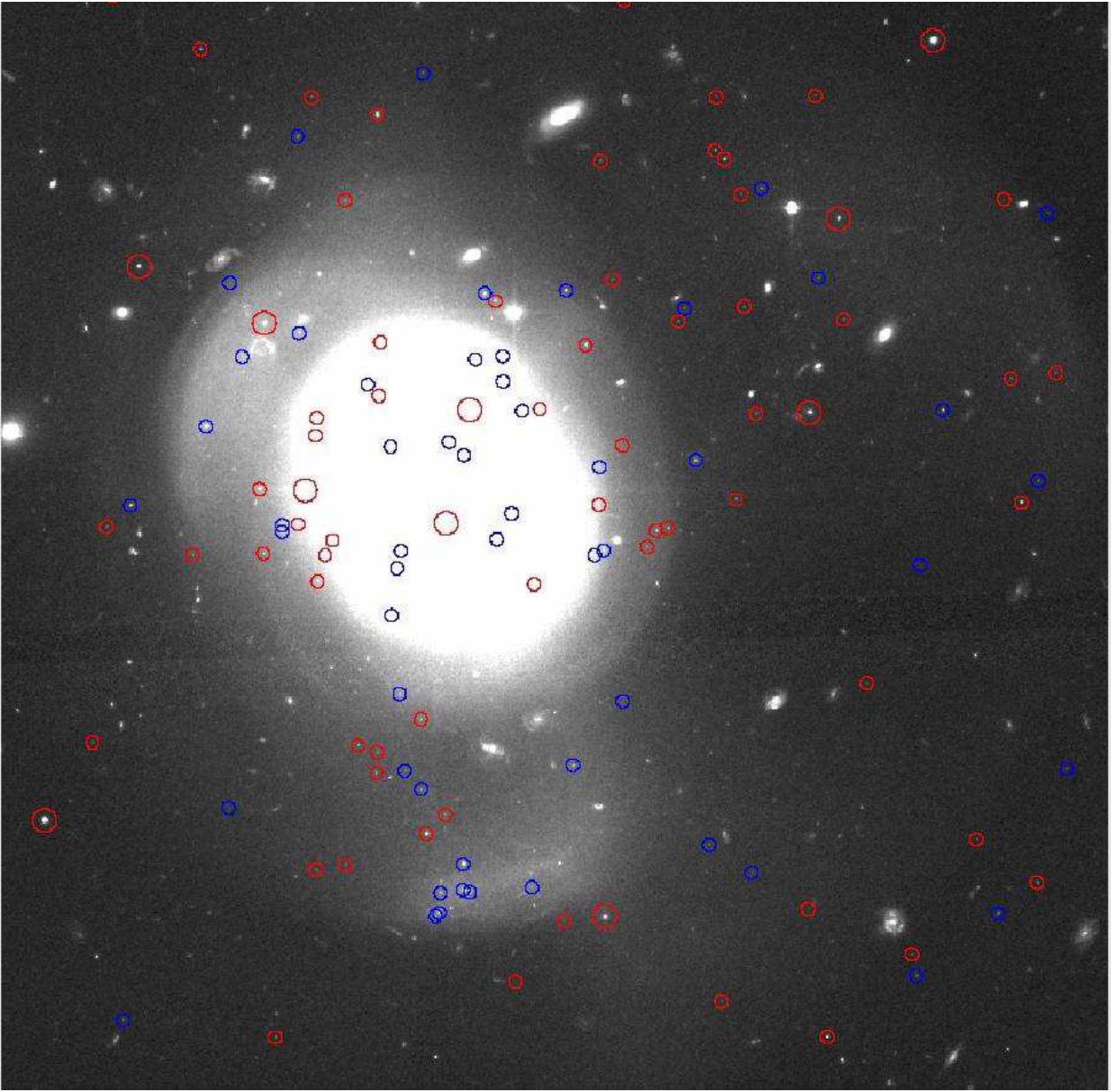}
\caption{AM 0139-655 through the F814W filter. The frame is 84\arcsec $\times$84\arcsec. The directions on the sky are same as in Fig. 1. The blue and the red circles represent the blue and the red clusters respectively. Clusters with {\it I} brighter than 24 mag are shown by the big red circles and clusters with {\it I} fainter than 24 mag are shown by the small red circles. Note that the bright shell to the East of the galaxy is populated exclusively by blue clusters. \label{clust}}
\end{figure}

Figure~\ref{figredblue} shows the projected spatial distribution of the blue and
the red clusters while Fig.~\ref{ring} shows the
\gI\ color map obtained by dividing the {\it g} image by the {\it I} image. In Fig.~\ref{ring}, note the presence of a blue ring-like structure surrounding the center
and the blue clusters present along the ring.
Another interesting fact is that the brightest and the most distinct shell
(lying between 14.8-16.7 kpc to the
East of the galaxy center) harbors seven blue clusters
(Fig.~\ref{clust}), 
all of which fall
within the color range 0.70 $\leq$ \gI\ $\leq$ 0.91, consistent with SSP model ages
between 0.25-0.4 Gyr. Thus, it appears that all the clusters associated
with this shell were formed during the latest burst of star formation. We
suggest that the same event that formed the shell structure also gave rise to
the young globular cluster population.

\section{Luminosity Function of AM 0139-655}
\begin{figure}
\plotone{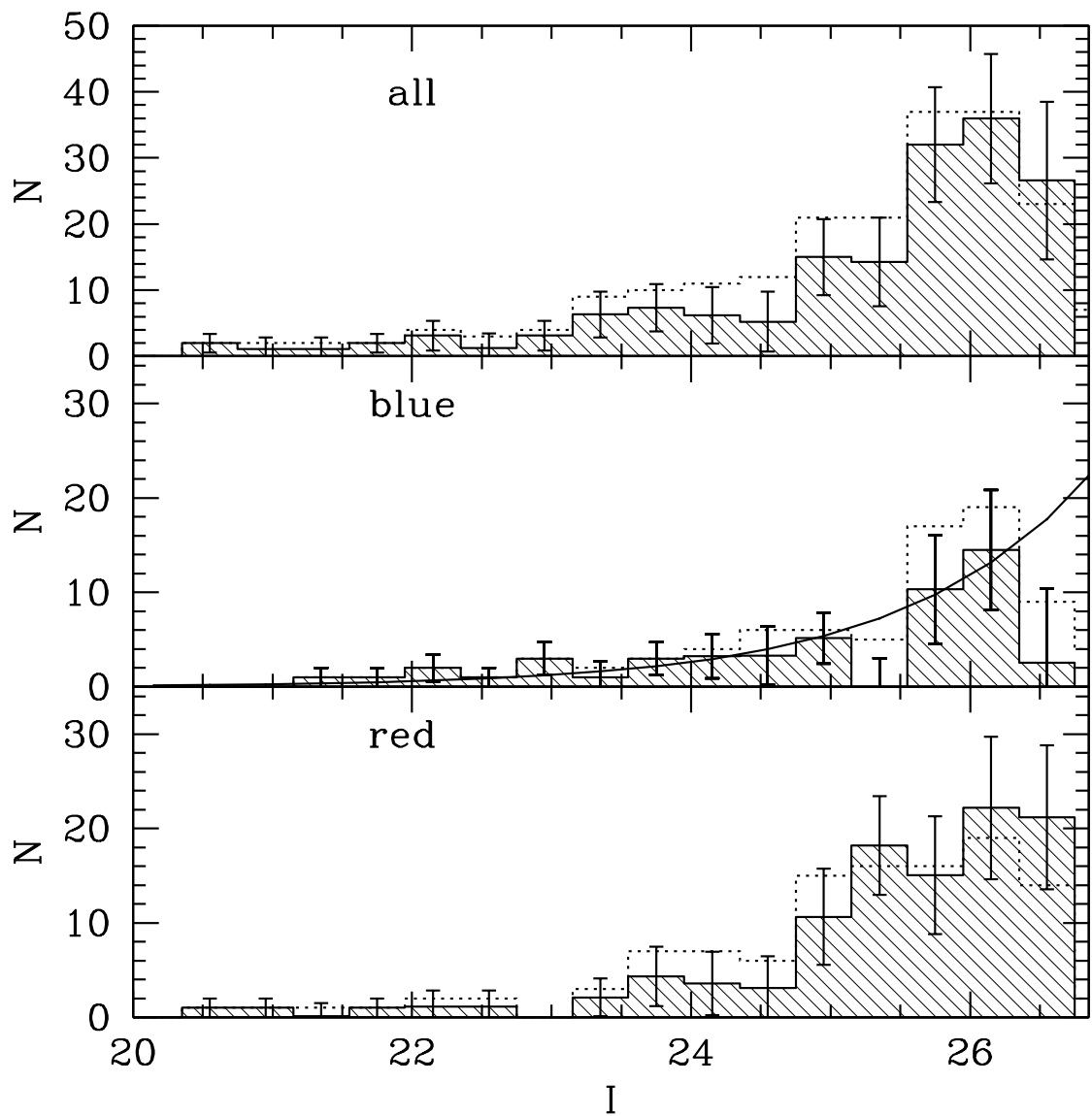}
\caption{The luminosity function of clusters in AM 0139-655 in the {\it I} band. The dotted histograms mark uncorrected (observed) LFs and the hatched histograms mark LFs corrected for completeness and background contamination.Completeness-weighted error bars are also plotted. The top panel shows the combined luminosity function of all the clusters. The middle panel shows the luminosity function of the blue clusters and the power law fit, giving an index $\alpha$ $\sim$ -1.8. The lower panel shows the luminosity function of the red clusters.}
\label{lf}
\end{figure}

To derive the luminosity functions (LFs) of the clusters, we proceeded in the following manner.
A completeness correction was applied to each
cluster depending on its magnitude and background level. This was determined
from the curves shown in Fig.~\ref{complete} using bilinear interpolation.
The clusters were then divided into two groups depending on their distance
from the galaxy center. The clusters within a projected distance of 40 kpc
from the center of the galaxy were considered as actual GC candidates and the
ones farther were considered as contaminants (foreground stars and compact
background galaxies). The LFs of the GC candidates were then corrected for
background contamination using the smoothed and scaled LFs of these objects.
The LFs obtained after completeness and background contamination correction
and the uncorrected LFs in the {\it I}-band are shown in Fig.~\ref{lf}.

Given the distance modulus of 35.3 for AM 0139-655, the
globular cluster turn-over magnitude for old, metal-poor GCs is expected at
{\it I} = 27.2 (for {\it $M_I$} = -8.1) \citep{harris96}. 
As seen from the figure, the data
does not quite reach the peak of the turnover. We analyze the
luminosity functions of the blue and red clusters separately.

\subsection{Luminosity Function of the Blue Clusters}
As mentioned above in Section 5.1, all clusters with colors in
the range
$0.0 \leq\ g-I \leq 0.95$ are considered
as blue clusters. Figure~\ref{lf} (middle panel) shows the luminosity function of
these blue clusters.
A power-law fit to the LF of the blue clusters yields
$\phi$(L)dL $\propto$ L$^{-\alpha}$ dL with $\alpha$ = $-1.8\pm0.1$,
consistent with the results of previous LF studies of young clusters
in mergers and merger remnants which find power-law indices in the range $-2.1 <  \alpha  <  -1.7$
\citep{whit1995,meurer1995,whit99,whit2003}.

\subsection{Luminosity Function of the Red Clusters}
Clusters within the color range $1.15 \leq\ g-I \leq 2.0$ were
classified as red clusters and their luminosity functions derived (Fig.~\ref{lf}, bottom panel).
Attempts to fit their luminosity function
by either a single gaussian or a single power law yielded large
residuals.
Upon close examination of the color-magnitude diagram, we found a significant number
of very luminous red clusters. If the red clusters were attributed solely to an
old, metal-poor population, with a mass function similar to that in our
Galaxy, the brightest clusters would not
be expected to be brighter than $M_g$ = $-$10.0 ({\it g} = 25.0), yet we find an excess of
clusters as bright as {\it g} = 23.5. In addition, the color distribution
of the red clusters
shows a broad tail towards red colors (Fig.~\ref{gi}) which suggests the
presence of more than one sub-population.

\subsubsection{Modeling the Composite Luminosity Function}

\begin{figure}
\plotone{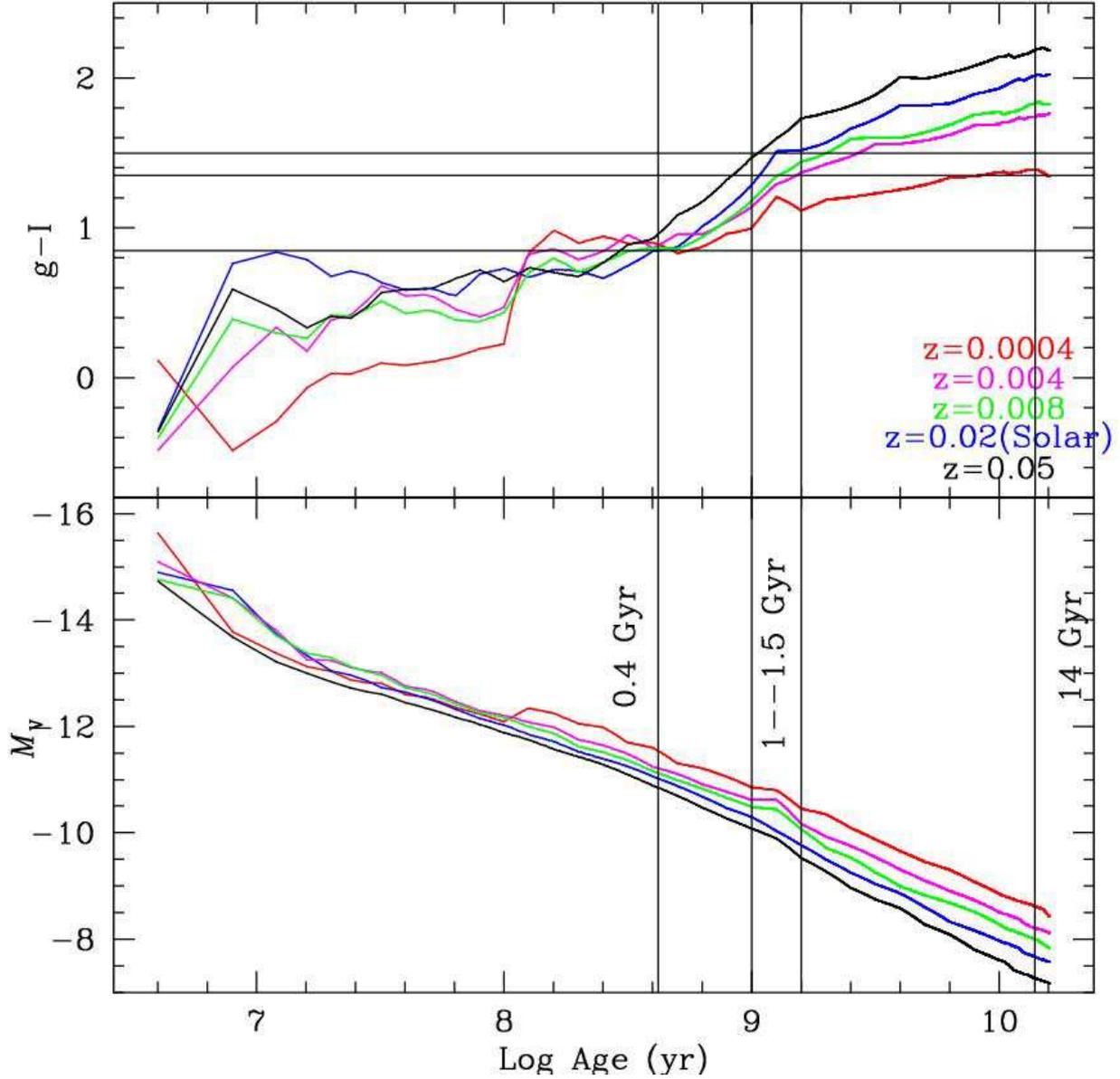}
\caption{Time evolution of the \gI\ color index (top panel) and absolute
magnitude {\it $M_V$} per 1$\times$ $10^{6}$ $M_{\odot}$ of single-burst stellar
population models using GALEV \citep{galev}. Model curves are plotted for a \citep{
sal} IMF and metallicities as indicated in the figure. Ages of the three stellar
 populations are indicated by 0.4 Gyr, 1-1.5 Gyr and 14 Gyr for the youngest,
the intermediate and the old population respectively. The evolution of the {\it V}
 magnitude, predicted from the same models is shown in the bottom panel and is
used to compute the fading of various clusters to an age of 14 Gyr.}
\label{time}
\end{figure}

Spectroscopic studies of intermediate-age clusters in NGC 1316 \citep{gmk} and NGC 5128 \citep{pff} have shown that
they have near-solar metallicities. Simulations of the color-magnitude diagrams
of evolving clusters by \citet{whitmore97} show that solar metallicity
clusters in the age
range of 1-2 Gyr are the most difficult to identify since they have nearly
the same colors as the old, metal-poor GC subpopulation.
However, they are significantly (2-2.5 mag) brighter than the old ones for a given mass.
\citet{whitmore97} also find that solar metallicity clusters older than $\sim$\, 2 Gyr will be
redder than the old, metal-poor cluster population.
Using the SSP models shown in Fig.~\ref{time}, we see that the range of colors corresponding to the red
cluster distribution in AM 0139-655 can indeed be explained by a metal-poor
population about 14 Gyr old, but also by a metal-rich younger population  about
1-1.5 Gyr old.
To test whether our observed distribution may indeed be the result of two
distinct populations of the same color but different ages and metallicities,
we model the observed luminosity function as a combination of two populations;
one
old, metal-poor population which is consistent with a gaussian magnitude
distribution and
the other an intermediate-age population of solar metallicity, which can be described
by a power law \citep{goud}.

To constrain the possible relative contributions of the two populations, we
consider the
specific frequency of GCs,
\begin{equation}
S_N = N_{GC}10^{0.4(M_V + 15)}
\end{equation}
i.e., the number of star clusters per galaxy luminosity normalized
to an absolute {\it V} magnitude of $-$15
\citep{harris81}. To estimate the contribution of the old,
metal-poor cluster population to the mixed luminosity function, we need to first consider the possible $S_N$
values of the progenitors and their {\it V} magnitude. Assuming AM 0139-655
was formed by a merger of two galaxies, the progenitor galaxies could have
been
two spirals or one elliptical and one spiral. Since there is no
a priori obvious way of determining the nature of the progenitors, we consider various
scenarios.

The specific frequency is known to increase systematically along the
Hubble sequence, from 0.5 $\pm$ 0.2 for Sc spirals to 2.6 $\pm$ 0.5 for
ellipticals outside of clusters \citep{harris}. Hence, we consider four values of
$S_N$ (0.25, 0.5, 1.0 and 3.0) covering the range from spirals to
ellipticals to describe the possible specific frequencies  of the progenitors. We use the
{\it V} magnitude of AM 0139-655 to estimate the combined {\it V} magnitude of the
progenitors. This will be a rough estimate and will tend to overestimate the
luminosity since star formation during the merger would brighten the merger
remnant.
However, the {\it V} magnitude of AM 0139-655
is not listed in the literature. Hence, we first calculate the total {\it g} and
{\it I} magnitudes of AM 0139-655 from our images using ellipse fits and
calculate \gI\ from these magnitudes. We get {\it g} = 13.64; {\it I} = 12.34 and hence, \gI\ = 1.30.

\begin{figure}
\plotone{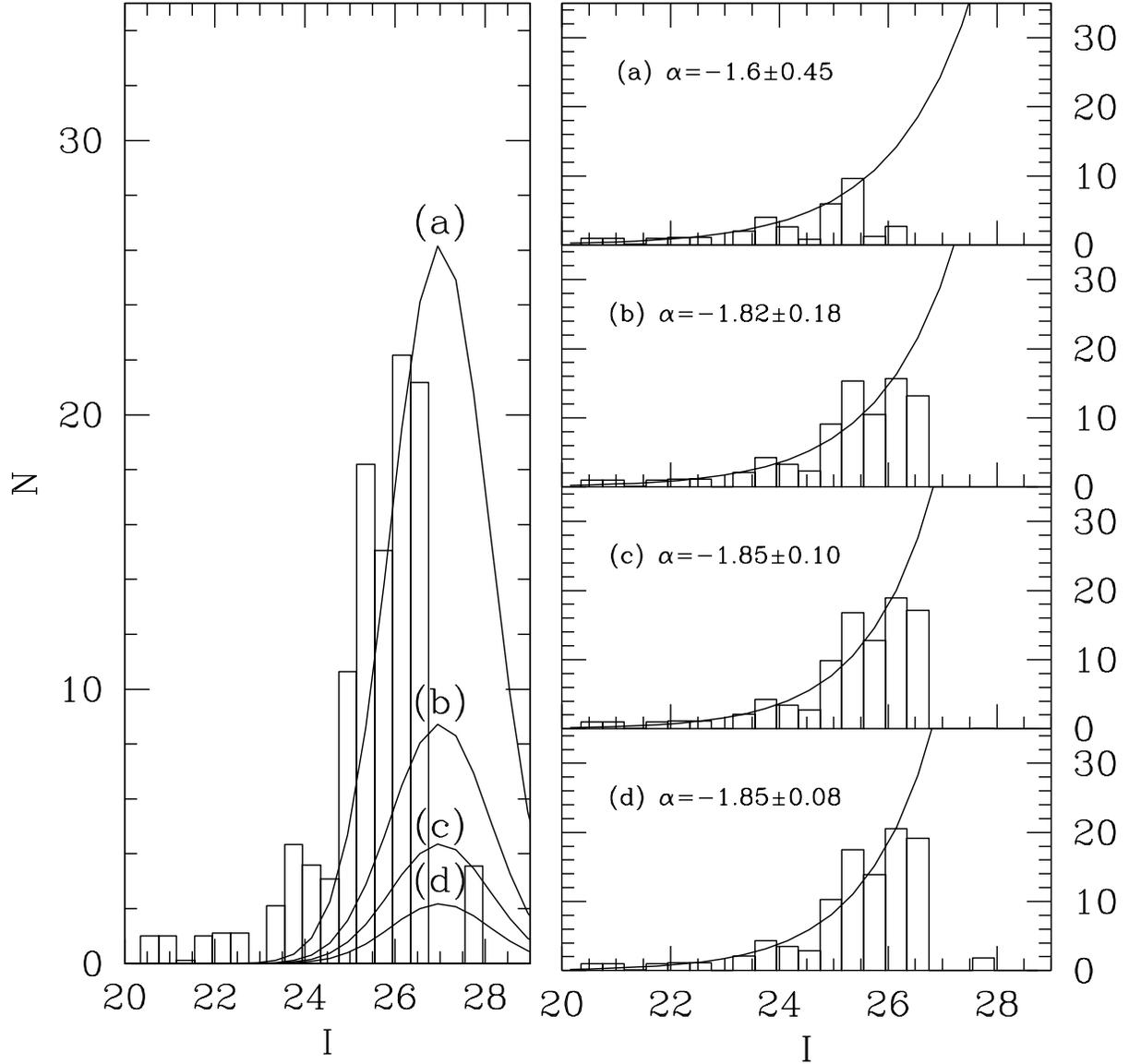}
\caption{The left panel is the luminosity function (histogram) of the red clusters. The curves a,b,c and d stand for estimated Gaussians representing the
old metal-poor GC population with specific frequencies of 3.0, 1.0, 0.5 and 0.25 respectively. The histograms in the right panel shows the luminosity functions of the clusters obtained after subtracting the contribution from each of the gaussians in the left panel. The solid lines in the right panels are power-law fits to the remaining clusters. Note that these clusters can be fit by a power-law (solid line) with $-1.6 \le \alpha \le -1.85$ as expected for young GCs. See Sect. 6.2 for a detailed discussion. \label{twopop}}
\end{figure}

Using this,
we find the luminosity-weighted age of 1 Gyr for a solar metallicity population at this \gI\ from Fig.~\ref{time}.
A 1 Gyr solar metallicity population has
$V\!-\!I$ = 0.89  and hence {\it V} = 13.23 according to the GALEV models.
We then compute the fading in the {\it V} band from this age to an age
of 14 Gyr from the lower panel of Fig.~\ref{time}.
We use luminosity-weighted age at solar metallicity as we are attempting
to separate the intermediate-age clusters (which should have about solar metallicity)
from the old, metal-poor GC population. Though a SSP is not a very good
approximation to a galaxy, it is sufficient for our purpose since we are
interested in getting a rough
estimate of the fading  of the galaxy.
Using the faded value of {\it V} = 15.84 ($M_V$ = -19.45), in the equation for $S_N$,
we calculate the number of old globular clusters expected for the
four values given above. The estimated gaussians for the old, metal-poor population were calculated using $M_I$ = $-$8.1 and $\sigma$ = 1.2 \citep{harris96, barmby2001} for each $S_N$.
Fig.~\ref{twopop} shows the observed luminosity function of the red clusters
and the estimated gaussians for the four values of
$S_N$. Subtracting the contributions to the gaussians in each case and
examining the residuals, we find that the residuals in each case can be fit
by power laws with power-law indices $\alpha$ between -1.6 and -1.85 consistent with values found
for intermediate-age cluster populations. The rms values of the
difference between the power law fits and the histograms in the right panel of
Fig.~\ref{twopop} are 4.08, 2.88, 3.28 and 3.02 for $S_N$ = 3.0, 1.0, 0.5 and 0.25 respectively.
The above analysis suggests that the red cluster population in AM 0139-655
is in fact a combination of two populations of different ages.
To conclude, we have identified three populations of GCs in AM 0139-655:
one blue, metal-rich population of young clusters and a red population that
can be split into two populations, one old (14 Gyr) metal-poor and one of
intermediate age (1-1.5 Gyr) and solar metallicity. Since the two
populations contributing to the red peak had similar colors, the KMM test
could not identify them as separate populations.
The estimates of ages,
metallicities and the number of clusters (corrected for background contamination
and completeness) belonging to the three sub-populations
are given in Table~\ref{tbl-3} for the various possible scenarios.
However, spectroscopic observations of the brightest candidates from
the three populations will be necessary to derive exact metallicities and
verify our tentative interpretation.

\subsection{Masses of the Clusters}
It is illustrative to estimate the masses of the bright clusters formed during
the merger and compare them to the masses of the most massive clusters in our
Galaxy ($\omega$ Cen with a mass $5 \times 10^{6} M_{\odot}$) and M31
(Mayall II, G1), with a mass $(1.4 - 1.7) \times 10^{7} M_{\odot}$  \citep{meylan2001}.

\begin{figure}
\plotone{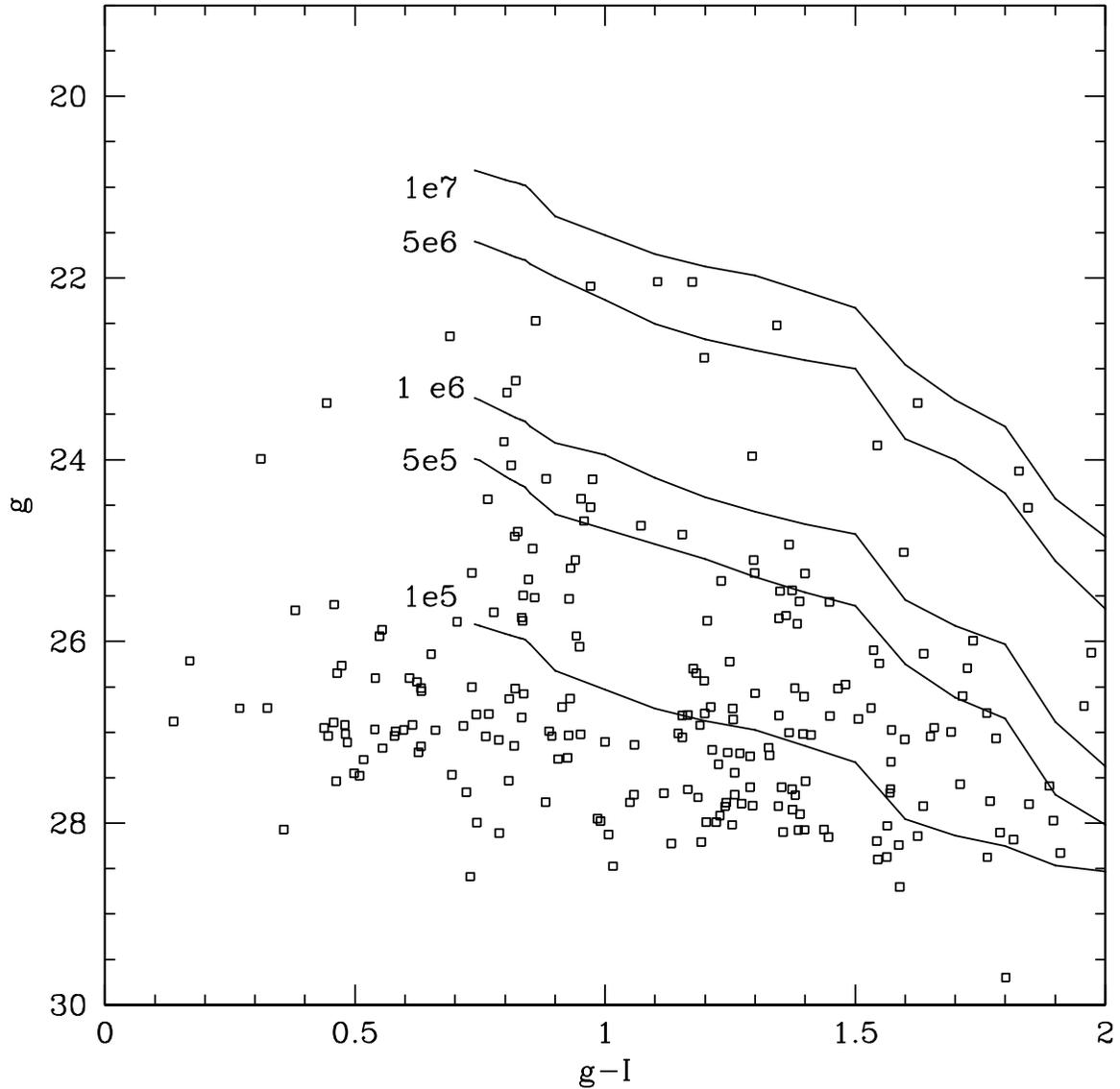}
\caption{The CMD of the clusters associated with AM 0139-655. Superimposed are curves of constant mass (in units of M$_\odot$). All curves are computed for solar metallicity with ages ranging from 0.3 Gyr at the leftmost point to 14 Gyr at the right extreme for each curve. The masses are the masses of individual clusters after fading them to 14 Gyr and which have current color and magnitudes as specified in the plot. Note that AM 0139-655 harbors several very massive red clusters. It is clear that several of these clusters will end up with masses equal to or greater than that of $\omega$ Cen.\label{massc}}
\end{figure}

Fig.~\ref{massc} shows curves of constant mass superimposed on
the color-magnitude diagram. To derive these masses, a grid was constructed
for the observed magnitude and color range. These magnitudes were then faded to an age of
14 Gyr. Masses were assigned to each cluster by assuming the value of $M/L_{V}$ to be
4.1 (at 14 Gyr), as found by \cite{meylan}. We find
that several clusters in AM 0139-655 will
have masses greater than that of the most massive cluster in our galaxy and
comparable to the most massive cluster G1 in M31 when they age to 14 Gyr
without subsequent mass loss due to stellar evolution since
this mass loss is small beyond 1 Gyr \citep{galev}. The clusters will lose
stars during their dynamical evolution. Their stellar-dynamical mass loss,
however, cannot be evaluated with any confidence without orbital information.

The masses derived in this way have two caveats. First, the mass-to-light ratio
is assumed to be equal to that of $\omega$ Cen. Various studies \citep{mdm}
find a wide range (1.2 to 5.2) of mass-to-light ratios for GCs. According to
\citet{bc03} models computed using the \citet{chabrier03} IMF and lower and
upper mass cut-offs $m_L = 0.1M_{\odot}$ and $m_U = 100M_{\odot}$, $M/L_V$
of a 14 Gyr solar metallicity population
is 4.3 while that of a $1/50\ Z_{\odot}$ population of the same age is 2.0. 
Hence, a mass-to-light ratio of 4.1 is a good approximation for solar
metallicity clusters. In the absence of
any spectroscopic data on these clusters, these values can be taken as rough
estimates. Secondly, the effects of internal reddening (though expected to be small from Sect. 5), have not been taken into
account. This is not expected to change the mass ranges much since
dereddening would tend to shift the points to brighter magnitudes and bluer
colors (upward and to the left in Fig.~\ref{massc}), which is nearly parallel to the curves of
equal mass.

\section{Nature of the Progenitors}
In a recent study of GC systems around six shell ellipticals, \citet{sik} typically
find bimodal distributions similar to those of normal ellipticals.
They find some indications of the presence of
a younger intermediate-age globular cluster population in two of their
galaxies, but the number of these intermediate-age clusters is small
as compared to the old clusters.
However, in the case of AM 0139-655, we find that the
cluster system is dominated by the young and the intermediate-age
clusters with a relatively small contribution from the old, metal-poor clusters.
In order to constrain the number of old, metal-rich clusters that might
have been inherited from the progenitors, we note that there is some
evidence for a small peak
which is consistent with a redder, metal-rich sub-population (at \gI\ = 1.8) 
as expected for a sub-solar metallicity from the SSP models. The small
number of GCs in this minor peak suggests
the absence of a significant bulge
in either precursor. Additionally, for the four values of $S_N$ considered
for modeling the LF of the precursors, the rms values point out that at least
one of the progenitors was not an elliptical, but had small $S_N$ values
$\leq$ 1, which is consistent with Hubble types of Sb or later \citep{rupali, goudfroo03}
which have small bulges. Both the above
findings seem to suggest that the galaxies involved in the merger
that formed AM 0139-655 did not have a large bulge generally seen in
early-type galaxies but
were most likely late-type spirals with small bulges. A merger involving
late-type gas-rich spirals could have formed one population of new GCs
about 1 Gyr ago in a starburst at first pericenter and another one 0.4 Gyr
ago at final merging \citep{mihos96}.

From the fact that we see evidence for a young and intermediate-age component
of clusters
in AM 0139-655, we can safely assume that at least one of the galaxies
involved in the recent interaction 
must have been rich in gas since there were two episodes of star formation within
the time periods for shell survival. This is also supported by the fact that AM 0139-655 itself has strong
Balmer absorption lines, which suggests that at least one of the progenitors
must have been gas-rich and hence could not have been an early-type with a
large bulge.

\section{Summary and Conclusions}

Deep ACS images of the post-starburst shell galaxy AM 0139-655 have been analyzed. We find a total of 217 GC candidates associated with this galaxy.
The \gI\ color distribution of the clusters shows two distinct peaks,
one at \gI\ = 0.85 and the other at \gI\ = 1.35. We compare the colors with SSP models and find that the red peak
coincides with the color expected of the old, metal-poor population. The
blue peak corresponds to a young population (0.4 Gyr) assuming solar metallicity.
The luminosity function of the blue clusters can be fit by a power law of
index $\alpha$ $\sim$ -1.8$\pm$0.1, as is expected for young clusters.
The red part of the distribution shows a long tail and the presence of several
luminous clusters.
We show that the luminosity function of the red clusters can be modeled as a 
combination of two subpopulations:
An old population with a gaussian distribution and
an intermediate-age population with a power law distribution.
Thus, three subpopulations of globular clusters are clearly distinguishable
in AM 0139-655. The number of clusters in the old, metal-rich subpopulation
that is commonly associated with ``normal" early-type galaxies is very small
relative to the three aforementioned subpopulations. The combination of this
dearth of old, metal-rich clusters and the fact that the LF of the old
population is best fit by a small value of $S_N$ supports the notion
that any bulge in the progenitor galaxies must have been small.

The ages of the two younger GC subpopulations are comparable to  
typical lifetimes of shell structures formed during simulations of  
galaxy mergers, and the brightest shell actually seems to host  
several young GCs. This strongly suggests that both the shells and  
the young and intermediate-age GCs were formed during a dissipative  
merger. It thus seems fairly likely that the E+A signatures of this  
galaxy are due to this recent merger. These findings lend credence to  
the ``merger scenario" for the formation of metal-rich GCs and their  
galaxy hosts.

\acknowledgments
We would like to thank the referee, Uta Fritze for useful comments.
We would like to thank Rupali Chandar and Diane Karakla for useful discussions.
Support for this work was provided by NASA through HST grant number GO-10227 from
the Space Telescope Science Institute, which is operated by the Association of
Universities for Research in Astronomy, Inc., under NASA contract NAS5-26555.


{\it Facilities:}  \facility{HST (ACS)}.

{}


\clearpage

\begin{deluxetable}{ll}
\tablecaption{General properties of AM 0139-655. \label{tbl-1}}
\tablewidth{0pt}
\tablehead{
\colhead{Parameter} & \colhead{Value}}
\startdata
RA$^*$(J2000)&1h 41m 30.98s\\
Dec$^*$(J2000)&$-$65$\degr$ 36$\arcmin$ 55.44$\arcsec$\\
Hubble Type&S0\\
${\it v}_{hel}$ &8216 $\pm$ 15 km s$^{-1}$\\
${\it v}_{hel}$ (corr)&7936 km s$^{-1}$\\
velocity dispersion&249 $\pm$ 46 km s$^{-1}$\\
Distance$^{**}$&105.8 Mpc\\
{\it m-M}&35.3\\
scale&1$\arcsec$ = 512 pc\\
&1 ACS pixel = 25pc\\
${\it M}_{B}$& $-$20.57\\
galactic latitude (b) & $-$50.7$\degr$\\
$A_v^{***}$&0.00\\
\enddata
\tablenotetext{*}{From ACS images, this work.}
\tablenotetext{**}{Using ${\it H}_{0}$ = 75 km s$^{-1}$Mpc$^{-1}$}
\tablenotetext{***}{\citet{burstein}}
\tablecomments{All other parameters are taken from LEDA (http://leda.univ-lyon1.fr/)}
\end{deluxetable}

\begin{deluxetable}{ccrrrr}
\tablewidth{0pt}
\tablecaption{Positions and photometry of the 50 brightest globular cluster candidates \label{tbl-2}}
\tablehead{
\colhead{RA (2000)}&\colhead{Dec (2000)}&
\colhead{{$I$} (mag)}&\colhead{{$g-I$} (mag)} &
\colhead{$r^{*}(\arcsec)$} & \colhead{$r^{**}$ (kpc)}}
\startdata

1 41 30.55 & -65 36 45.99 & 20.87 $ \pm $ 0.01 & 1.17 $ \pm $ 0.02 & 9.82 & 5.0 \\
1 41 30.03 & -65 35 54.43 & 20.94 $ \pm $ 0.01 & 1.11 $ \pm $ 0.02 & 61.29 & 31.4 \\
1 41 31.13 & -65 36 51.80 & 21.12 $ \pm $ 0.01 & 0.97 $ \pm $ 0.02 & 3.77 & 1.9 \\
1 41 27.32 & -65 37 41.27 & 21.18 $ \pm $ 0.01 & 1.34 $ \pm $ 0.02 & 51.10 & 26.1 \\
1 41 31.79 & -65 36 51.03 & 21.61 $ \pm $ 0.01 & 0.86 $ \pm $ 0.02 & 6.68 & 3.4 \\
1 41 33.59 & -65 36 20.00 & 21.68 $ \pm $ 0.01 & 1.20 $ \pm $ 0.02 & 38.96 & 19.9 \\
1 41 31.42 & -65 36 55.59 & 21.75 $ \pm $ 0.03 & 1.62 $ \pm $ 0.07 & 2.72 & 1.4 \\
1 41 41.69 & -65 37 15.45 & 21.95 $ \pm $ 0.01 & 0.69 $ \pm $ 0.02 & 69.30 & 35.5 \\
1 41 28.43 & -65 36 46.53 & 22.30 $ \pm $ 0.01 & 1.83 $ \pm $ 0.02 & 18.14 & 9.3 \\
1 41 30.32 & -65 36 53.15 & 22.31 $ \pm $ 0.02 & 0.82 $ \pm $ 0.02 & 4.69 & 2.4 \\
1 41 24.96 & -65 36 20.35 & 22.46 $ \pm $ 0.01 & 0.80 $ \pm $ 0.02 & 51.20 & 26.2 \\
1 41 24.19 & -65 37 29.70 & 22.67 $ \pm $ 0.01 & 1.29 $ \pm $ 0.02 & 54.23 & 27.7 \\
1 41 21.20 & -65 37 32.10 & 22.68 $ \pm $ 0.01 & 1.85 $ \pm $ 0.02 & 70.79 & 36.2 \\
1 41 30.46 & -65 36 57.50 & 22.93 $ \pm $ 0.02 & 0.44 $ \pm $ 0.03 & 3.82 & 1.9 \\
1 41 29.46 & -65 36 40.09 & 23.00 $ \pm $ 0.02 & 0.80 $ \pm $ 0.02 & 18.01 & 9.2 \\
1 41 31.12 & -65 36 52.59 & 23.24 $ \pm $ 0.03 & 0.98 $ \pm $ 0.04 & 2.98 & 1.5 \\
1 41 31.78 & -65 36 58.93 & 23.25 $ \pm $ 0.02 & 0.81 $ \pm $ 0.03 & 6.05 & 3.1 \\
1 41 29.62 & -65 37 3.23 & 23.33 $ \pm $ 0.02 & 0.88 $ \pm $ 0.02 & 11.44 & 5.8 \\
1 41 31.33 & -65 37 24.40 & 23.42 $ \pm $ 0.02 & 1.60 $ \pm $ 0.03 & 29.03 & 14.8 \\
1 41 26.70 & -65 36 25.75 & 23.48 $ \pm $ 0.02 & 0.95 $ \pm $ 0.03 & 39.80 & 20.4 \\
1 41 24.35 & -65 36 23.94 & 23.55 $ \pm $ 0.02 & 0.97 $ \pm $ 0.03 & 51.72 & 26.5 \\
1 41 21.14 & -65 37 18.86 & 23.57 $ \pm $ 0.02 & 1.37 $ \pm $ 0.03 & 65.24 & 33.4 \\
1 41 26.62 & -65 36 5.87 & 23.66 $ \pm $ 0.02 & 1.07 $ \pm $ 0.03 & 56.46 & 28.9 \\
1 41 27.33 & -65 36 38.57 & 23.67 $ \pm $ 0.02 & 1.15 $ \pm $ 0.03 & 28.21 & 14.4 \\
1 41 35.83 & -65 36 49.06 & 23.67 $ \pm $ 0.02 & 0.77 $ \pm $ 0.03 & 30.74 & 15.7 \\
1 41 30.67 & -65 36 58.31 & 23.68 $ \pm $ 0.04 & 0.31 $ \pm $ 0.05 & 3.45 & 1.7 \\
1 41 30.16 & -65 36 54.38 & 23.72 $ \pm $ 0.03 & 0.96 $ \pm $ 0.04 & 5.20 & 2.6 \\
1 41 29.12 & -65 37 30.60 & 23.81 $ \pm $ 0.02 & 1.30 $ \pm $ 0.03 & 36.99 & 18.9 \\
1 41 30.15 & -65 36 59.70 & 23.85 $ \pm $ 0.02 & 1.40 $ \pm $ 0.04 & 6.68 & 3.4 \\
1 41 36.63 & -65 36 58.83 & 23.95 $ \pm $ 0.02 & 1.30 $ \pm $ 0.03 & 35.15 & 18.0 \\
1 41 36.02 & -65 36 46.84 & 23.97 $ \pm $ 0.02 & 0.83 $ \pm $ 0.03 & 32.40 & 16.6 \\
1 41 35.53 & -65 36 49.60 & 24.02 $ \pm $ 0.02 & 0.82 $ \pm $ 0.03 & 28.78 & 14.7 \\
1 41 26.32 & -65 36 59.17 & 24.06 $ \pm $ 0.02 & 1.37 $ \pm $ 0.03 & 29.06 & 14.9 \\
1 41 35.04 & -65 36 47.56 & 24.10 $ \pm $ 0.02 & 1.35 $ \pm $ 0.03 & 26.35 & 13.5 \\
1 41 21.85 & -65 36 2.19 & 24.10 $ \pm $ 0.02 & 1.23 $ \pm $ 0.03 & 77.69 & 39.8 \\
1 41 29.50 & -65 36 52.78 & 24.12 $ \pm $ 0.03 & 0.85 $ \pm $ 0.03 & 9.51 & 4.8 \\
1 41 22.49 & -65 37 31.88 & 24.15 $ \pm $ 0.02 & 1.97 $ \pm $ 0.04 & 63.96 & 32.7 \\
1 41 32.33 & -65 36 49.64 & 24.16 $ \pm $ 0.02 & 0.94 $ \pm $ 0.03 & 10.18 & 5.2 \\
1 41 29.73 & -65 36 48.36 & 24.17 $ \pm $ 0.03 & 1.39 $ \pm $ 0.04 & 10.49 & 5.4 \\
1 41 42.40 & -65 37 29.54 & 24.26 $ \pm $ 0.02 & 1.74 $ \pm $ 0.04 & 78.53 & 40.2 \\
1 41 31.52 & -65 37 0.55 & 24.26 $ \pm $ 0.03 & 0.93 $ \pm $ 0.04 & 6.11 & 3.1 \\
1 41 33.11 & -65 37 37.96 & 24.35 $ \pm $ 0.03 & 1.36 $ \pm $ 0.04 & 44.52 & 22.8 \\
1 41 24.02 & -65 36 42.34 & 24.40 $ \pm $ 0.03 & 1.35 $ \pm $ 0.04 & 45.02 & 23.0 \\
1 41 29.77 & -65 37 9.49 & 24.42 $ \pm $ 0.03 & 1.38 $ \pm $ 0.04 & 15.90 & 8.1 \\
1 41 29.56 & -65 37 1.17 & 24.47 $ \pm $ 0.03 & 0.85 $ \pm $ 0.04 & 10.45 & 5.3 \\
1 41 20.68 & -65 36 45.31 & 24.50 $ \pm $ 0.03 & 1.64 $ \pm $ 0.04 & 64.55 & 33.0 \\
1 41 28.81 & -65 37 3.23 & 24.51 $ \pm $ 0.03 & 0.73 $ \pm $ 0.04 & 15.49 & 7.9 \\
1 41 38.81 & -65 37 12.53 & 24.56 $ \pm $ 0.03 & 1.54 $ \pm $ 0.04 & 51.41 & 26.3 \\
1 41 34.14 & -65 35 44.84 & 24.57 $ \pm $ 0.03 & 1.20 $ \pm $ 0.04 & 73.27 & 37.5 \\
1 41 24.77 & -65 36 31.55 & 24.57 $ \pm $ 0.03 & 1.72 $ \pm $ 0.05 & 45.29 & 23.2 \\

\enddata
\tablenotetext{*}{projected radius derived using the position of the center at RA = 1h 41m 30.98s and Dec = -65$\degr$ 36$\arcmin$ 55.44$\arcsec$}
\tablenotetext{**}{Using ${\it H}_{0}$ = 75 km s$^{-1}$Mpc$^{-1}$}
\end{deluxetable}

\clearpage

\begin{table}
\caption{Properties of the three sub-populations of GCs}
\label{tbl-3}
\begin{tabular}{cccc}
\tableline\tableline
&Age (Gyr)&Z&N\\
\tableline
Old&14&0.0004&25 (for $S_N=1.00$)\\
&&&12 (for $S_N=0.50$)\\
&&&6 (for $S_N=0.25$)\\
Intermediate&1$--$1.5&0.02&81 (for $S_N=1.00$)\\
&&&93 (for $S_N=0.50$)\\
&&&99 (for $S_N=0.25$)\\
Young&0.4&0.02&50\\
\tableline
\end{tabular}
\end{table}
\clearpage

\begin{thebibliography}{}
\bibitem[Anders \& Fritze-v. Alvensleben(2003)]{galev} Anders, P., \& Fritze-v. Alvensleben, U. 2003, \aap, 401, 1063
\bibitem[Arp(1966)]{arp} Arp, H. 1966, \apjs, 123, 1
\bibitem[Ashman \& Zepf(1992)]{ash92} Ashman, K.M., \& Zepf, S.E. 1992, \apj, 384, 50
\bibitem[Ashman et al.(1994)]{abz} Ashman, K.M., Bird, C.M., \& Zepf, S.E. 1994, \aj, 108, 2348
\bibitem[Ashman \& Zepf(1998, and references therein)]{ash98} Ashman, K.M., \& Zepf, S.E. 1998, In {\it Globular Cluster Systems}, (Cambridge Univ. Press, Cambridge)
\bibitem[Barmby, Huchra, \& Brodie(2001)]{barmby2001} Barmby, P., Huchra, J.P., \& Brodie, J, P. 2001, \aj, 121, 1482
\bibitem[Bruzual \& Charlot(1993)]{bru} Bruzual, G. A., \& Charlot, S. 1993, \apj, 405, 538
\bibitem[Bruzual \& Charlot(2003)]{bc03} Bruzual, G. A., \& Charlot, S. 2003, \mnras, 344, 1000
\bibitem[Burstein \& Heiles(1982)]{burstein} Burstein, D., \& Heiles, C. 1982, \aj, 87, 1165
\bibitem[Carter et al.(1988)]{carter} Carter, D., Prieur, J.L., Wilkinson, A., Sparks, W.B., \& Malin, D.F. 1988, \mnras, 235, 813
\bibitem[Chabrier(2003)]{chabrier03} Chabrier, G. 2003, \pasp, 115, 763
\bibitem[Chandar et al.(2006)]{chandar} Chandar, R., Puzia, T.H., Sarajedini, A., \& Goudfrooij, P. 2006, \apj, 646, 107
\bibitem[Chandar et al.(2004)] {rupali} Chandar, R., Whitmore, B., \& Lee, M.G. 2004, \apj, 611, 220
\bibitem[Charmandaris et al.(2000)]{char} Charmandaris, V., Combes, F., \& van der Hulst, J.M. 2000, \aap, 35, L1
\bibitem[Cohen, Blakeslee, \& C\^{o}te(2003)]{cbc03} Cohen, J.G., Blakeslee, J.P., \& C\^{o}te, P. 2003, \apj, 592, 866
\bibitem[C\^{o}te, Marzke, \& West(1998)]{cmw} C\^{o}te, P., Marzke, R.O., \& West, M.J. 1998, \apj, 501, 554
\bibitem[Dressler \& Gunn(1983)]{dressler83} Dressler, A., \& Gunn, J.E. 1983, \apj, 270, 7
\bibitem[Dupraz \& Combes(1986)]{dc86} Dupraz, C., \& Combes, F. 1986, \aap, 166, 53
\bibitem[Fabian et al.(1980)]{fabian} Fabian, A.C., Nulsen, P.E.J., \& Stewart, G.C. 1980, \nat, 287, 613
\bibitem[Forbes et al.(2001)]{forbes01} Forbes, D.A., Beasley, M.A., Brodie, J.P., \& Kissler-Patig, M. 2001, \apj, 563, L143
\bibitem[Forbes, Brodie, \& Grillmair(1997)]{forbes} Forbes, D. A., Brodie, J.P., \& Grillmair, C.J. 1997, \aj, 113, 1652
\bibitem[Forte, Martinez, \& Muzzio(1982)]{fmm} Forte, J.C., Martinez, R.E., \& Muzzio, J.C. 1982, \aj, 87, 1465
\bibitem[Franx(1993)]{franx93} Franx, M. 1993, \pasp, 105, 1058
\bibitem[Goudfrooij et al.(2001)]{gmk} Goudfrooij, P., Mack, J., Kissler-Patig, M., Meylan, G., \& Minniti, D. 2001, \mnras, 322, 643
\bibitem[Goudfrooij et al.(2003)]{goudfroo03} Goudfrooij, P., Strader, J., Brenneman, L., Kissler-Patig, M., Minniti, D., \& Huizinga, E.J. 2003, \mnras, 343, 665
\bibitem[e.g. Goudfrooij et al.(2004)]{goud} Goudfrooij, P, Gilmore, D., Whitmore, B.C., Schweizer, F. 2004, \apj, 613, L121
\bibitem[Goudfrooij et al.(2007)]{gnew} Goudfrooij, P., Schweizer, F., Gilmore, D., \& Whitmore, B.C. 2007, {\it astro-ph/0702467}
\bibitem[Harris \& van den Bergh(1981)]{harris81} Harris, W.E., \& van den Bergh, S. 1981, \aj, 86, 1627
\bibitem[Harris(1991)]{harris} Harris, W.E. 1991, \araa, 29, 543
\bibitem[Harris(1996)]{harris96} Harris, W.E. 1996, \aj, 112, 1487
\bibitem[Hernquist \& Quinn(1987)]{hq87} Hernquist, L., \& Quinn, P.J. 1987, \apj, 312, 1
\bibitem[Hernquist \& Quinn(1988)]{hq88} Hernquist, L., \& Quinn, P.J. 1988, \apj, 331, 682
\bibitem[Hernquist \& Quinn(1989)]{hq89} Hernquist, L., \& Quinn, P.J. 1989, \apj, 342, 1
\bibitem[Jord\'{a}n et al.(2005)]{jordan} Jord\'{a}n, A., C\^{o}te, P., Blakeslee, J. P.. et al. 2005, \apj, 634, 1002
\bibitem[Koekemoer et al.(2002)]{kok} Koekemoer, A.M., Fruchter, A.S., Hook, R.N., \& Hack, W. 2002, in {\it The 2002 HST Calibration Workshop: Hubble after the Installation of the ACS and the NICMOS Cooling System}, ed. S. Arribas, A. Koekemoer, \& B. Whitmore (Baltimore: Space Telescope Science Institute), p.337
\bibitem[Kojima \& Noguchi(1997)]{kn} Kojima, M., \& Noguchi, M. 1997, \apj, 481, 132
\bibitem[Kundu \& Whitmore(2001)]{kundu} Kundu, A., \& Whitmore, B. C. 2001, \aj, 121, 2950
\bibitem[Kundu et al.(2005)]{kundu05} Kundu, A., Zepf, S.E., Hempel, M., et al. 2005, \apj, 634, L41
\bibitem[Malin \& Carter(1983)]{malin} Malin, D.F., \& Carter, D. 1983, \apj, 274, 534
\bibitem[McLachlan \& Basford(1988)]{mb88} McLachlan, G. J., \& Basford, K. E. 1988, In:{\it Mixture Models: Inference and Application to Clustering} (New York: Dekker)
\bibitem[Meurer et al.(1995)]{meurer1995} Meurer, G.R., Heckman, T.M., Leitherer, C., Kinney, A., Robert, C., \& Garnett, D.R. 1995, \aj, 110, 2665
\bibitem[Meylan et al.(1991)]{mdm} Meylan, G., Dubath, P., \& Mayor, A. 1991, ASPC, 13, 158
\bibitem[Meylan et al.(1995)]{meylan} Meylan, G., Mayor, M., Duquennoy, A., \& Dubath, P. 1995, \aap, 303, 761
\bibitem[Meylan et al.(2001)]{meylan2001} Meylan, G., Sarajedini, A., Jablonka, P., Djorgovski, S.G., Bridges, T., \& Rich, R.M. 2001, \aj, 122, 830
\bibitem[Mihos \& Hernquist(1996)]{mihos96} Mihos, J.C., \& Hernquist, L. 1996, \apj, 464, 641
\bibitem[Miller et al.(1997)]{mwsf} Miller, B.W., Whitmore, B.C., Schweizer, F., \& Fall, S.M. 1997, \aj. 114, 2381
\bibitem[Norton et al.(2001)]{norton} Norton, S.A., Gebhardt, K., Zabludoff, A.I., \& Zaritsky, D. 2001, \apj, 557, 150
\bibitem[Nulsen(1989)]{nulsen89} Nulsen, P.E.J. 1989, \apj, 346, 690
\bibitem[Peng et al.(2004)]{pff} Peng, E.W., Ford, H.C., \& Freeman, K.C. 2004, \apj, 602, 705
\bibitem[Prieur(1990)]{prieur} Prieur, J.-L. 1990, In: {\it Dynamics and Interactions of Galaxies}, ed. R. Wielen, Springer-Verlag Berlin, p. 72
\bibitem[Puzia et al.(2002)]{puzia02} Puzia, T.H., Zepf, S.E., Kissler-Patig, M., Hilker, M., Minniti, D., \& Goudfrooij, P. 2002, \aap, 391, 453
\bibitem[Puzia et al.(2005)]{puzia05} Puzia, T.H., Kissler-Patig, M., Thomas, D., Maraston, C., Saglia, R.P., Bender, R., Goudfrooij, P., \& Hempel, M. 2005, \aap, 439, 997
\bibitem[Quinn(1984)]{quinn1984} Quinn, P. 1984, \apj, 279, 596
\bibitem[Salpeter(1955)]{sal} Salpeter, E.E. 1955, \apj, 121, 161
\bibitem[Schiminovich et al.(1994)]{sch} Schiminovich, D., van Gorkom, J.H., van der Hulst, J.M., \& Kasow, S. 1994, \apj, 423, 101
\bibitem[Schweizer(1980)]{schweizer1980} Schweizer, F. 1980, \apj, 237, 303
\bibitem[Schweizer \& Ford(1985)]{sf} Schweizer, F., \& Ford, W.K. 1985, in {\it New Aspects of Galaxy Photometry}, ed. J.-L. Nieto, New York:Springer-Verlag, p. 145
\bibitem[Schweizer(1987)]{schweizer87} Schweizer, F. 1987, In: {\it Nearly Normal Galaxies} ed. by S.M. Faber (Springer, New York) p.18
\bibitem[Schweizer et al.(1996)]{smwf} Schweizer, F., Miller, B.W., Whitmore, B.C., \& Fall, S.M. 1996, \aj, 112, 1839
\bibitem[Schweizer \& Seitzer(1998)]{ss} Schweizer, F., \& Seitzer, P. 1998, \aj, 116, 2206
\bibitem[Schweizer et al.(2004)]{ssb} Schweizer, F., Seitzer, P., \& Brodie, J.P. 2004, \aj, 128, 202
\bibitem[Sikkema et al.(2006)]{sik} Sikkema, G., Peletier, R.F., Carter, D., Valentijn, E.A., \& Balcells, M. 2006, \aap, 458, 53
\bibitem[Silverman(1986)]{silverman86} Silverman, B.W. 1986, In: {\it Density Estimation for Statistics and Data Analysis}, Chap and Hall/CRC Press, Inc.
\bibitem[Sirianni et al.(2005)]{siri05} Sirianni, M., et al. 2005, \pasp, 117,1409
\bibitem[Stetson(1987)]{stet} Stetson, P.B. 1987, \pasp, 113, 1420
\bibitem[e.g. Terlevich et al.(1999)]{terlevich} Terlevich, A.I., Kuntschner, H., Bower, R.G., Caldwell, N., \& Sharples, R.M.  1999, \mnras, 310, 445
\bibitem[van den Bergh \& Mackey(2004)]{vanden} van den Bergh, S., \& Mackey, A. D. 2004, \mnras, 354, 713
\bibitem[Whitmore \& Schweizer(1995)]{whit1995} Whitmore, B.C., \& Schweizer, F. 1995, \aj, 109, 960
\bibitem[Whitmore et al.(1997)]{whitmore97} Whitmore, B.C., Miller, B.W., Schweizer, F., \& Fall, S.M. 1997, \aj, 114, 1797
\bibitem[Whitmore et al.(1999)]{whit99} Whitmore, B.C., Zhang, Q., Leitherer, C., Fall, S.M., Schweizer, F., \& Miller, B.W. 1999, \aj, 118, 1551
\bibitem[Whitmore(2003)]{whit2003} Whitmore, B.C. : The Formation of Star Clusters, In: {\it A Decade of HST Science} ed. by M.Livio et al. (Cambridge Univ. Press, Cambridge 2003) p.153
\bibitem[Williams \& Christiansen(1985)]{wc} Williams, R.E., \& Christiansen, W.A. 1985, \apj, 291, 80
\bibitem[Zabludoff et al.(1996)]{zab96} Zabludoff, A.I., Zaritsky, D., Lin, H., Tucker, D., Hashimoto, Y., Shectman, S.A., Oemler, A., \& Kirshner, R.P. 1996, \apj, 466, 104

\end{thebibliography}
\end{document}